%% file: PIP105_LATESTVERSION.tex
\documentclass[traditabstract, longauth]{aa} 
\usepackage{graphicx}
\usepackage{float}
\usepackage{txfonts}
\usepackage[breaklinks, colorlinks, citecolor=blue]{hyperref}
\usepackage{natbib}

\bibpunct{(}{)}{;}{a}{}{,} 

\newcommand{\nilc}{{\tt NILC}}
\newcommand{\sevem}{{\tt SEVEM}}
\newcommand{\smica}{{\tt SMICA}}
\newcommand{\commander}{{\tt COMMANDER}}

\newcommand{\wmap}{\rm WMAP}

\newcommand{\cluster}{{\tt CLUSTER}}
\newcommand{\galaxy}{{\tt GALAXY}}

\newcommand{\vlosr}{v_{\rm los}^{\rm rec}}
\newcommand{\vloso}{v_{\rm los}^{\rm orig}}
\newcommand{\vn}{\hat{\mbox{\vec{n}}}}

\newcommand{\vx}{\vec{x}}

\newcommand{\vk}{\vec{k}}

\newcommand{\vrv}{\vec{r}}

\newcommand{\vv}{\vec{v}}

\newcommand{\gsim}{\; ^{>}_{\sim}\;}

\newcommand{\be}{\begin{equation}}
\newcommand{\ee}{\end{equation}}

\usepackage{multirow}

\providecommand{\sorthelp}[1]{}

\input Planck.tex

\begin{document}
%
\input{PIP_105_Hernandez_authors_and_institutes.tex}
  \title{\Planck\ intermediate results. XXXVII. Evidence of unbound gas from the kinetic Sunyaev-Zeldovich effect.}
\authorrunning{\Planck\ Collaboration}  
\titlerunning{kSZ evidence from unbound gas}

\date{\today}

\abstract{By looking at the kinetic Sunyaev-Zeldovich effect (kSZ) in \Planck\ nominal mission data, we present a significant detection of baryons participating in large-scale bulk flows around central galaxies (CGs) at redshift $z\approx 0.1$.
We estimate the pairwise momentum of the kSZ temperature fluctuations at the positions of the Central Galaxy Catalogue (CGC) samples extracted from Sloan Digital Sky Survey (SDSS-DR7) data. For the foreground-cleaned \sevem, \smica, \nilc, and \commander\ maps, we find $1.8$--$2.5\,\sigma$ detections of the kSZ signal, which are consistent with the kSZ evidence found in individual \Planck\ raw frequency maps, although lower than found in the \wmap-9yr W band ($3.3\,\sigma$).  We further reconstruct the peculiar velocity field from the CG density field, and compute for the first time the cross-correlation function between kSZ temperature fluctuations and estimates of CG radial peculiar velocities. This correlation function yields a $3.0$--$3.7$\,$\sigma$ detection of the peculiar motion of extended gas on Mpc scales in flows correlated up to distances of 80--100\,$h^{-1}$\,Mpc. Both the pairwise momentum estimates and the kSZ temperature-velocity field correlation find evidence for kSZ signatures out to apertures of 8\,arcmin and beyond, corresponding to a physical radius of $> 1\,$Mpc, more than twice the mean virial radius of halos. This is consistent with the predictions from hydrodynamical simulations that most of the baryons are outside the virialized halos. We fit a simple model, in which the temperature-velocity cross-correlation is
proportional to the signal seen in a semi-analytic model built upon {\it N}-body
simulations, and interpret the proportionality constant as an effective optical
depth to Thomson scattering.  We find  $\tau_{\rm T}=(1.4\pm0.5)\times 10^{-4}$; the
simplest interpretation of this measurement is that much of the gas is in a diffuse
phase, which contributes little signal to X-ray or thermal Sunyaev-Zeldovich observations.
}



\authorrunning{Planck Collaboration}

\maketitle

\section{Introduction}

The kinetic Sunyaev-Zeldovich effect \citep[hereafter kSZ,][]{kSZ0,kSZ} describes the Doppler boost experienced by a small fraction of the photon bath of the cosmic microwave background (CMB) radiation when scattering off a cloud of moving electrons. 
In the limit of Thomson scattering, where there is no energy exchange between the electrons and the CMB photons, the kSZ effect is equally efficient for all frequencies and gives rise to relative brightness temperature fluctuations in the CMB that are frequency independent. The exact expression for this effect was first written as equation~15 of  \citet[][]{kSZveryfirst}, and reads
\begin{equation}
\frac{\delta T}{T_{\rm 0}} (\vn) = -\int {\rm d}l\,\sigma_{\rm T} n_{\rm e} \left(\frac{\vv}{c} \cdot \vn\right).
\label{eq:kSZ1}
\end{equation}
In this expression, the integral is performed along the line of sight, $n_{\rm e}$ denotes the electron number density, $\sigma_{\rm T}$ is the Thomson cross-section, and $(\vv/c)\cdot \vn$ represents the line of sight component of the electron peculiar velocity in units of the speed of light $c$. The equation above shows that the kSZ effect is sensitive to the peculiar {\em momentum} of the free electrons, since it is proportional to both their density and peculiar velocity. A few previous studies \citep{chm05,Bhattacharya08DE,MaZhao13,ks_pksz13,Liangulowhite2014} have proposed that it be used to trace the growth of velocities throughout cosmic history and its connection to dark energy and modified gravity. \citet{Zhang_homog} and \citet{planck2013-XIII} have also used the kSZ effect to test the Copernican Principle and the homogeneity of the Universe. In addition, the impact of the kSZ effect on sub-cluster scales has also been investigated \citep{inogamov_ras, kdolag_ras13}.

With these motivations, there have been previous attempts to detect the kSZ effect in existing CMB data \citep[][]{kashlinsky08,kashlinsky10,lavaux_afshordi12,Handetal2012}. The results of \citet[][]{kashlinsky08,kashlinsky10}, which point to the existence of a bulk flow of large amplitude (800--1000\,km\,s$^{-1}$) extending to scales of  at least 800 Mpc, have been disputed by a considerable number of authors \citep[e.g.,][]{keisler09,osborneetal11,modyandamir,planck2013-XIII,snbf2013}.  On the other hand, \citet[][]{lavaux_afshordi12} have claimed the detection of the local bulk flow (within $80\,h^{-1}$\,Mpc) by using the kSZ in WMAP data in the direction of nearby galaxies.  While these results remain at a low (roughly $1.7\,\sigma$) significance level, they are also in slight tension with the results that we present here. On the other hand, the work of \citet[][]{Handetal2012} constitutes the first clear detection of the kSZ effect (using the ``pairwise'' momentum approach that we describe below), and no other group has confirmed their results to date. In addition, \citet{kSZcso} have provided a first claimed detection of the kSZ effect in a single source. After the first data release of the \Planck\ mission, some disagreement has been claimed between the cosmological frame set by \Planck\ and measurements of peculiar velocities as inferred from redshift space distortions \citep[][]{macaulay13}. More recently, \citealt{Mueller2014} forecasted the detectability of the neutrino mass with the kSZ pairwise momentum estimator by using Atacama Cosmology Telescope (ACTPol) and Baryon Oscillation Spectroscopic Survey (BOSS) data.
 
This work represents the second contribution of the \Planck\
\footnote{\Planck\ (\url{http://www.esa.int/Planck}) is a project of the 
European Space Agency  (ESA) with instruments provided by two scientific 
consortia funded by ESA member states and led by Principal Investigators 
from France and Italy, telescope reflectors provided through a collaboration 
between ESA and a scientific consortium led and funded by Denmark, and 
additional contributions from NASA (USA).}
 collaboration to the study of the kSZ effect, in which we focus on constraining the ``missing baryons'' with this signal \citep{dedeoetal05,chm08,hoetalkSZ09,chm09,kSZtomography1}. Numerical simulations~\citep{cenostriker2006} have shown that most of the baryons lie outside galactic halos, and in a diffuse phase, with temperature in range of $10^{5}$--$10^{7}$ K. This ``warm-hot'' intergalactic medium is hard to detect in X-ray observations due to the relatively low temperature. Thus, most of the baryons are ``missing" in the sense that they are neither hot enough ($T<10^{8}$~K) to be seen in X-ray observations, nor cold enough ($T>10^{3}$~K) to be made into stars and galaxies. However, as we can see in Eq.~(\ref{eq:kSZ1}), the kSZ signal is proportional to the gas density and peculiar velocity, and thus the gas temperature is irrelevant.  Therefore the kSZ effect has been proposed as a ``leptometer," since it is sensitive to the ionized gas in the Universe while most of the baryons remain undetected.

In this work, we measure the kSZ effect through two distinct statistics: the kSZ pairwise momentum, which was used in \citet{Handetal2012} to detect kSZ effect for the first time; and the cross-correlation between the kSZ temperature fluctuations with the reconstructed radial peculiar velocities inferred from a galaxy catalogue. These analyses allow us to probe the amount of gas generating the kSZ signal and provide  direct evidence of the elusive missing baryons in the local Universe. In Sect.~\ref{sec:datades} we describe the CMB data and the Central Galaxy Catalogue that will be used in our analysis. In Sect.~\ref{sec:method} we describe two statistical tools we use. Section~\ref{sec:results} presents and explains our results and illustrates the physical meaning of the significance. The conclusions and discussion is presented in the last section. Throughout this work we adopt the cosmological parameters consistent with \citet{planck2014-a15}: $\Omega_{\rm m} = 0.309$; $\Omega_{\Lambda}=0.691$; $n_{\rm s}=0.9608$; $\sigma_{8}=0.809$; and $h=0.68$, where the Hubble constant is $H_0=100\,h$\,km\,s$^{-1}$\,Mpc$^{-1}$.


\section{Data description}
\label{sec:datades}
\subsection{\Planck\ data}

This work uses \Planck\ data that are available publicly, both raw frequency maps and the CMB foreground cleaned maps.\footnote{\Planck's Legacy Archive:  \\\url{http://pla.esac.esa.int/pla/aio/planckProducts.html}} The kSZ effect should give rise to frequency-independent temperature fluctuations and hence constitutes a secondary effect with identical spectral behaviour as for the intrinsic CMB anisotropies. Therefore the kSZ effect should be present in all CMB frequency channels and in all CMB foreground-cleaned maps. However, one must take into account the different effective angular resolutions of each band, particularly when searching for a typically small-scale signal such as the kSZ effect. The raw frequency maps used here are the LFI 70\,GHz map, and the HFI 100, 143, and 217\,GHz maps. These have effective FWHM values of 13.01, 9.88, 7.18, and 4.87\,arcmin for the 70, 100, 143, and 217\,GHz maps, respectively.

The FWHM of the foreground cleaned products is  5\,arcmin for the maps used in this work, namely  the \sevem, \smica, \nilc, and \commander\ maps. These maps are the output of four different component-separation algorithms. While the \nilc\ map is the result of an internal linear combination technique, \smica\ uses a spectral matching approach, \sevem\ a template-fitting method and \commander\ a parametric, pixel based Monte Carlo Markov Chain technique to project out foregrounds. We refer to the \Planck\ component separation papers for details in the production of these maps \citep{planck2014-a11,planck2014-a12}. In passing, we note that the foreground subtraction in those maps is not perfect, and that there exist traces of residuals, particularly on the smallest angular scales, related to radio, dust, and thermal Sunyaev-Zel'dovich [tSZ] emission.

For comparison purposes, we examine 9-year data from the {\it Wilkinson} Microwave Anisotropy Probe (\wmap)\footnote{\wmap: \url{http://map.gsfc.nasa.gov}}. In particular, we downloaded from the LAMBDA site\footnote{LAMBDA: \url{http://lambda.gsfc.nasa.gov}} the \wmap\ satellite 9-year W-band (94\,GHz) map, with an effective FWHM of 12.4\,arcmin.

\subsection{Central Galaxy Catalogue}
\label{sec:cgc}

We define a galaxy sample in an attempt to trace the centres of dark matter halos. Using as a starting point the seventh data release of the Sloan Digital Sky Survey \citep[SDSS/DR7][]{Abazajianetal2009}, the Central Galaxy Catalogue (CGC) is composed of 262\,673 spectroscopic sources brighter than $r=17.7$ (the $r$-band extinction-corrected Petrosian magnitude). These sources were extracted from the SDSS/DR7 New York University Value Added Galaxy Catalogue \citep{Blanton2005} \footnote{NYU-VAGC: \url{ http://sdss.physics.nyu.edu/vagc/ }}, and we have applied the following isolation criterion: no brighter (in $r$-band) galaxies are found within 1.0\,Mpc in the transverse direction and with a redshift difference smaller than 1\,000\,km\,s$^{-1}$. The Sloan photometric sample has been used to remove all possible non-spectroscopic sources that may violate the isolation requirements. By using the ``photometric redshift 2" catalogue \citep[photoz2,][]{Cunhaetal2009}, all spectroscopic sources in our isolated sample with potential photometric companions within 1.0\,Mpc in projected distance, and with more than 10\,\% probability of having a smaller redshift than that of the spectroscopic object, are dropped from the Central Galaxy Catalogue. 

Information provided in the NYU-VAGC yields estimates of the stellar mass content in this sample, and this enables us to make a direct comparison to the output of numerical simulations \citep{planck2012-XI}. We refer to this paper for further details on the scaling between total halo mass and stellar mass for this galaxy sample. 

We expect most of our CGs to be the central galaxies of their dark matter halos \citep[again see][]{planck2012-XI}, just as bright field galaxies lie at the centres of their satellite systems and cD galaxies lie near the centres of their clusters. They are normally the brightest galaxies in their system. By applying the same isolation criteria to a mock galaxy catalogue based on the Munich semi-analytic galaxy formation model (see Sec.~\ref{sec:sims} for more details), we found that at stellar masses  above  $10^{11}$\,M$_\odot$, more than 83\,\% of CGs in the mock galaxy catalogue are truly central galaxies. For those CGs that are satellites, we have checked that at $\log( M_\ast/\rm{M}_\odot)>11$ about two-thirds are brighter than the true central galaxies of their halos, while the remainder are fainter, but are considered isolated because they are more than 1\,Mpc (transverse direction) from their central galaxies (60\,\%) or have redshifts differing by more than 1\,000 \,km\,s$^{-1}$ (40\,\%). 

A case with stricter isolation criteria has been tested, in which we require no brighter galaxies to be found within 2.0\,Mpc in the transverse direction and with a redshift difference smaller than 2\,000\,km\,s$^{-1}$. Applying these stricter criteria to the same parent SDSS spectroscopic catalogue, we end up with a total of 110\,437 CGs, out of which which 58\,105 galaxies are more massive than $10^{11}$\,M$_\odot$. Thus about 30\,\% of the galaxies with $\log (\rm{M}_\ast/\rm{M}_\odot )>11$ have been eliminated from the sample. This new sample of CGs has a slightly higher fraction of centrals, reaching about 87\,\% at $\log (\rm{M}_\ast/\rm{M}_\odot)>11$. The improvement is small because (as we have checked) most of the satellite galaxies in the 1\,Mpc sample are brighter than the central galaxies of their own halos, and with the stricter criteria they are still included. Considering a balance between the sample size, which directly affects the signal-to-noise ratio in our measurements, and the purity of central galaxies, we choose the 1\,Mpc isolation criteria and the corresponding CGC sample in our analysis from here on. This sample amounts to 262\,673 sources over the DR7 footprint (about 6\,300\,deg$^2$, $f_{\rm sky}=0.15$).


\subsection{Numerical simulations}
\label{sec:sims}

In order to compare measurements derived from observations to theoretical
predictions, we make use of two different numerical simulations
of the large-scale-structure. We first look at the combination 
of two hydrodynamical simulations, combining a constrained 
realization of the local Universe and a large cosmological
simulation covering 1200\,$h^{-1}$\,Mpc. The simulations were performed
with the {\sc{GADGET-3}} code \citep{2001ApJ...549..681S,2005MNRAS.364.1105S}, 
which makes use of the entropy-conserving formulation of smoothed-particle hydrodynamics (SPH)
 \citep{2002MNRAS.333..649S}. 
These simulations include radiative cooling, heating by a uniform
redshift--dependent UV background \citep{1996ApJ...461...20H}, and a treatment of
star formation and feedback processes. The latter is based on a sub-resolution
model for the multiphase structure of the interstellar medium 
\citep{2003MNRAS.339..289S} with parameters that have been fixed to obtain
a wind velocity of around $ 350$\,km\,s$^{-1}$. We used the code {\tt SMAC} \citep{dolag2005}
to produce full sky maps from the simulations. For the innermost shell (up to 90\,$h^{-1}$\,Mpc from the observer) we use the simulation of the local Universe, whereas the rest of the shells
are taken from the large, cosmological box. We construct full sky maps 
of the thermal and kinetic SZ signals, as well as halo catalogues, by stacking
these consecutive shells through the cosmological boxes taken at the evolution
time corresponding to their distance. In this way, the simulations reach out to
a redshift of 0.22 and contain 13\,058 objects with masses above $10^{14}\,$M$_{\odot}$.
Provided that our CGC lies at a median redshift of 0.12, this combined simulation should be 
able to help us interpret the clustering properties of peculiar velocities of highly biased, massive halos, on the largest scales. For more details on the procedure for using the simulations see \citet{kdolag_ras13}. This simulated
catalogue will hereafter be referred to as the \cluster\ catalogue.

We have also used a mock galaxy catalogue, based on the semi-analytic galaxy formation simulation of \cite{2013MNRAS.428.1351G}, which is  implemented on the very large dark matter Millennium simulation \citep{2005Natur.435..629S}. The Millennium simulation follows the evolution of cosmic structure within a box of side 500\,$h^{-1}$\,Mpc (comoving), whose merger trees are complete for subhalos above 
a mass resolution of $1.7 \times 10^{10}$\,$h^{-1}$\,M$_\odot$. Galaxies are assigned to dark matter halos following the model recipes described in \cite{2011MNRAS.413..101G}. The rescaling technique of \cite{2010MNRAS.405..143A} has been adopted to convert the Millennium simulation,  which is originally based on \wmap-1 cosmology, to the \wmap-7 cosmology. The galaxy formation parameters have been adjusted to fit several statistical observables, such as the luminosity, stellar mass, and correlation functions for galaxies at $z=0$. 

We project the simulation box along one axis, and assign every galaxy a redshift based on its line of sight (LOS) distance and peculiar velocity, i.e., parallel to the LOS axis. In this way we can select a sample of galaxies from the simulation using criteria exactly analogous to those used for our CGC based on SDSS. A galaxy is selected if it has no brighter companions within 1\,Mpc in projected distance and 1\ 000\,km\,s$^{-1}$ along the LOS. This sample of simulated galaxies will be referred as the \galaxy\ catalogue.

\section{Methodology}
\label{sec:method}
We will search for kSZ signatures in \Planck\ temperature maps by implementing two different statistics on the CMB data. The first statistic aims to extract the kSZ pairwise momentum by following the approach of \citet{Handetal2012}, which was inspired by \citet{groth89} and \citet{roman98}. The second statistic correlates the kSZ temperature anisotropies estimated from \Planck\ temperature maps with estimates of the radial peculiar velocities. These velocity estimates are obtained after inverting the continuity equation relating galaxy density with peculiar velocities, as suggested by \citet{dedeoetal05} and \citet{hoetalkSZ09}. The particular inversion methods applied to our data are described in \citet[][hereafter K12]{shuandangulo12}. Throughout this work we  use the {\tt HEALPix} software \citep[][]{healpix}\footnote{\url{http://healpix.jpl.nasa.gov}} to deal with the CMB maps.

\subsection{The pairwise kSZ momentum estimator}
\citet{ferreiraetal99} developed an estimator for pairwise momentum with weights depending only on line-of-sight quantities. With this motivation we use a similar weighting scheme on our CMB maps.
The pairwise momentum estimator combines information on the relative spatial distance of pairs of galaxies with their kSZ estimates in a statistic that is sensitive to the gravitational infall of those objects. Specifically we use
\begin{equation}
\hat{p}_{\rm kSZ} (r) = -\frac{\sum_{i<j}(\delta T_{ i} - \delta T_{ j} )\,c_{ i,j}}{\sum_{ i<j} c_{ i,j}^2},
\label{eq:pksz1}
\end{equation}
where the weights $c_{ i,j}$ are given by 
\begin{equation}
c_{ i,j} = \hat{\vrv}_{ i,j} \cdot \frac{\hat{\vrv}_{ i}+\hat{\vrv}_{ j}}{2} = 
 \frac{(r_{ i}-r_{ j})(1+\cos\theta)}{2\sqrt{r_{ i}^2 + r_{ j}^2 - 2r_{ i}r_{ j}\cos\theta}}.
\label{eq:cweight}
\end{equation}
Here ${\vrv}_{ i}$ and ${\vrv}_{ j}$ are the vectors pointing to the positions of the $i$th and $j$th galaxies on the celestial sphere, $r_{ i}$ and $r_{ j}$ are the comoving distances to those objects, and ${\vrv}_{ i,j} = \vrv_{ i}-\vrv_{ j}$ is the distance vector for this galaxy pair. The {\it hat} symbol ($\hat{\vrv}$) denotes a unit vector in the direction of $\vrv$, and $\theta$ is the angle separating $\hat{\vrv}_{ i}$ and $\hat{\vrv}_{ j}$. We also note that $\hat{\vrv}_{i,j}$ means $(\vrv_i-\vrv_j)/|\vrv_i-\vrv_j|$. The sum is over all galaxy pairs lying a distance $r_{ i,j}$ falling in the distance bin assigned to $r$. The quantity $\delta T_{ i}$ denotes a {\em relative} kSZ temperature estimate at the position of the $i$th galaxy:
\begin{equation}
\delta T_{ i} = T_{\rm AP}(\vn_{ i}) - \bar{T}_{\rm AP}(z_i,\sigma_{ z}).
\label{eq:dT}
\end{equation}
In this equation, the symbol $T_{\rm AP}(\vn_{ i})$ corresponds to the kSZ amplitude estimate obtained at the angular position of the $i$th galaxy with an aperture photometry (AP) filter, while $\bar{T}_{\rm AP}(z_i,\sigma_{ z})$ denotes the average kSZ estimate obtained from {\em all} galaxies after weighting by a Gaussian  of width $\sigma_{ z}$ centred on $z_{ i}$:
\begin{equation}
\bar{T}_{\rm AP}(z_i,\sigma_{ z}) = 
 \frac{\sum_{ j} {T_{\rm AP}(\vn_{ j}) \exp{(-\frac{( z_{ i}-z_{ j} )^2}{2\sigma^2_{ z}} } )}}{\sum_{ j} \exp{(-\frac{( z_{ i}-z_{ j} )^2}{2\sigma^2_{ z}} )} }.
\label{eq:Tav}
\end{equation}
In this AP approach one computes the average temperature within a given angular radius $\theta$, and subtracts from it the average temperature in a surrounding ring of inner and outer radii $[\theta, f\theta]$, with $f>1$. In our case we use $f=\sqrt{2}$ \citep[see, e.g., ][]{planck2013-XIII}. If we knew accurately the spatial gas distribution on these scales there would clearly be room for more optimal approaches, like a matched filter technique \citep{Liangulowhite2014}. But on these relatively small scales the density profile in halos seems to be significantly less regular than the pressure profile \citep[see, e.g.,][]{arnaudetal10}, 
and this adds to the fact that gas outside halos, not necessarily following any given profile, also contributes to the signal. Furthermore, we would like to conduct an analysis that is as model independent as possible, and thus blind to any specific model for the gas spatial distribution.

The correction by a redshift-averaged quantity introduced in the last equation above was justified by \citet{Handetal2012} through the need to correct for possible redshift evolution of the tSZ signal in the sources. In practice, this correction allows us to measure {\em relative} changes in temperature anisotropies between nearby galaxy pairs after minimizing other noise sources, such as CMB residuals. 

\begin{figure}
\centering
\includegraphics[width=9.cm]{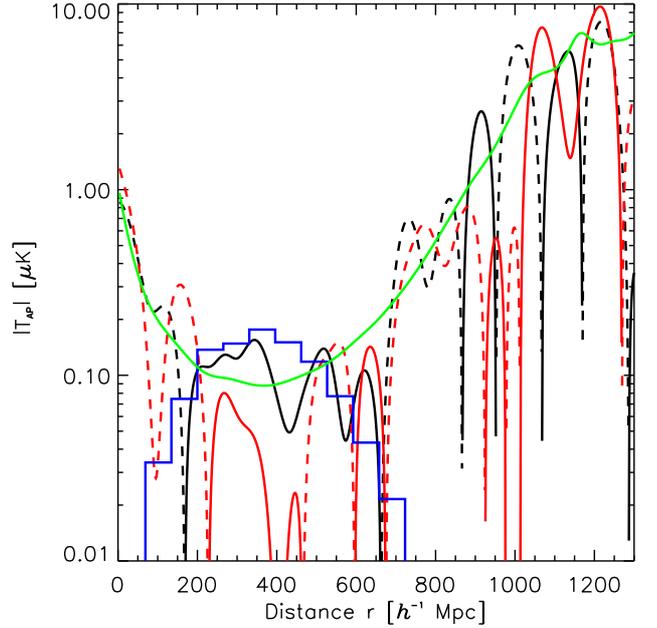}
\caption[fig:Tapmono]{ Absolute value of $\bar{T}_{\rm AP}(z(r),\sigma_{ z}) $ versus distance to the observer after obtaining AP kSZ temperature estimates on the real positions of CGs (black line) and on a rotated CG configuration (red line). The green line provides the theoretical expectation for the rms of $\bar{T}_{\rm AP}(z(r),\sigma_{ z})$, and the blue histogram displays the radial distribution of the CGs (see text for further details). }
\label{fig:Tapmono}
\end{figure}

We next conduct an analysis into the motivation behind this $z$-dependent monopole correction. In 
Fig.~\ref{fig:Tapmono} we display the amplitude of $| \bar{T}_{\rm AP}(z(r),\sigma_{ z})|$ for $\sigma_z=0.01$ versus the comoving distance to the observer $r$. The value of the aperture chosen in this exercise is 8\,arcmin. The black line refers to the real position of the CGs, while the red line refers to a rotated-on-the-sky configuration of the CGs. Since we are plotting absolute values of $\bar{T}_{\rm AP}(z,\sigma_{ z})$, the solid (dashed) parts of these lines refer to positive (negative) values of $\bar{T}_{\rm AP}(z,\sigma_{ z})$. The amplitude of these curves is close to the solid green line, which corresponds to the theoretical prediction for the rms of $\bar{T}_{\rm AP}(z,\sigma_{ z})$ under the assumption that the AP temperature estimates are dominated by CMB residuals. That is, the amplitude of the green curve equals $\sigma [\delta T] / \sqrt{N_{\rm CG}(r)}$, where $\sigma [\delta T] \simeq 20\,\mu$K is the rms of the 8\,arcmin AP kSZ estimates and $N_{\rm CG}(r)$ is the number of CGs effectively falling under the redshift Gaussian window given in Eq.~(\ref{eq:Tav}). While $\sigma [\delta T]\simeq 20\,\mu$K is computed from real data, its amplitude is very close to the theoretical predictions of Fig.~6 in \citet{chm05}, for an angular aperture of 8\,arcmin after considering the CMB exclusively.  We can see that the amplitude of  $| \bar{T}_{\rm AP}(z,\sigma_{ z})|$ is lowest for those distances where the number density of CGs is highest (as displayed, in arbitrary units, by the blue histogram in Fig.~\ref{fig:Tapmono}).

Therefore Fig.~\ref{fig:Tapmono} shows not only that AP kSZ estimates are dominated by CMB residuals, but also that the difference in $\bar{T}_{\rm AP}(z,\sigma_{ z})$ for CG pair members lying 50--100\,$h^{-1}$\,Mpc away in radial distance typically amounts to few times $0.01\,\mu$K. We shall show below that this amplitude is not completely negligible when compared to the typical kSZ pairwise momentum amplitude between CGs at large distances, and thus subtracting $\bar{T}_{\rm AP}(z,\sigma_{ z})$ becomes necessary. However, this should not be the case when cross-correlating AP kSZ measurements with estimates of radial peculiar velocities (see Sect.~\ref{sec:invertrho} below), since in this case the residual $\bar{T}_{\rm AP}(z,\sigma_{ z})$ will not be correlated with those velocities and should not contribute to this cross-correlation. 

Like \citet{Handetal2012}, our choice for $\sigma_z$ is $\sigma_z=0.01$, although results are very similar if we change this by a factor of 2. Adopting higher values introduces larger errors in the peculiar momentum estimates, and smaller values tend to suppress the power on the largest scales; we adopt $\sigma_z=0.01$ as a compromise value. It is worth noting that any effect giving rise to a pair of $\delta T_i, \delta T_j$, whose difference is not correlated to the relative distance of the galaxies $r_i-r_j$, should not contribute to Eq.~(\ref{eq:pksz1}).

\subsection{Recovering peculiar velocities from the CGC galaxy survey}
\label{sec:invertrho}

There is a long history of the use of the density field to generate estimates of peculiar velocities \citep[e.g.,][for some early studies]{dekel93,nusserdavis94,fisherlahavhoffman95,zaroubietal95}.
\citet{dedeoetal05} and \citet{hoetalkSZ09} first suggested inverting the galaxy density field into its peculiar velocity field in the context of kSZ studies. In this section, we use the approach of K12 to obtain estimates of the peculiar velocity field from a matter density tracer.  The work of K12 is based upon the analysis of the Millennium simulation, where full access to all dark matter particles was possible. In our case the analysis will obviously be limited to the use of the CGC catalogue, and this unavoidably impacts the performance of the algorithms. 

Inspired by K12, in our analysis we conduct three different approaches to obtain the velocity field.  The first one is a simple inversion of the linear continuity equation of the density contrast obtained from the galaxies in the CGC: this will be hereafter called the LINEAR approach. It makes use of the continuity equation at linear order,
\begin{equation}
\frac{\partial \delta (\vx)}{\partial t} + \nabla\vv (\vx ) = 0,
\label{eq:conteq}
\end{equation}
where $\vv (\vx )$ is the peculiar velocity field and $\delta (\vx )$ is the matter density contrast. In our case, however, we observe the galaxy density contrast $\delta^{\rm g} (\vx )$, and thus the amplitude of the estimated velocity field is modulated by the bias factor $b$ relating $\delta^{\rm g} (\vx )$ and $\delta (\vx)$,  $\delta^{\rm g} (\vx) = b \delta (\vx)$. The bias factor $b$ is assumed to be constant on the scales of interest.

The second approach performs the same inversion of the linear continuity equation, but on a {\em linearized} estimate of the density field. This linearized field is obtained after computing the natural logarithm of unity plus the density contrast of the galaxy number density field and subsequently removing its spatial average \citep[][]{neynricketal09},
\begin{equation}
\delta^{\rm LOG} (\vx ) = \ln{\left(1+ \delta^{\rm g} (\vx)\right) } - \langle \ln{\left(1+ \delta^{\rm g}\right) }\rangle_{\rm spatial}.
\label{eq:loglin1}
\end{equation}
In this expression, $\delta^{\rm g}(\vx)$ denotes again the density contrast of the galaxy number density at position $\vx$. This approach will be referred to as LOG-LINEAR. Finally, second-order perturbation theory \citep[see, e.g.,][]{bouchetetal95} was applied on this linearized field, yielding a third estimate of the peculiar velocity field (the LOG-2LPT approach). We refer to K12 for details on the implementations of the three approaches.  For the sake of simplicity, we discuss results for the LINEAR approach, and leave the corresponding discussion of the two other methods for an appendix. We have also tried using the full SDSS spectroscopic sample (rather than the CGC) on the same volume to recover the peculiar velocity field; the full  galaxy sample should be a better proxy for the dark matter density field than the CGC. However, after testing this with an enlarged version of the \galaxy\ catalogue (by considering all halos above a mass threshold of $10^{10.8}$\,M$_{\odot}$), we obtain negligible differences with respect to the mock CG. Likewise, we obtain practically the same results when using the real CGC and the full spectroscopic sample. Therefore, for the sake of simplicity, we restrict our analysis to the CGC. All these methods make use of FFTs requiring the use of a 3D spatial grid when computing galaxy number densities. We choose to use a grid of 128$^3$ cells, each cell being 4\,$h^{-1}$\,Mpc on a side. This cell size is well below the scales where the typical velocity correlations are expected (above 40\,$h^{-1}$\,Mpc), and comparable with the positional shifts induced by the redshift space distortions (about 3\,$h^{-1}$\,Mpc for a radial velocity of 300\,km\,s$^{-1}$). 

These distortions will be ignored hereafter, since they affect scales much smaller than those of interest in our study (roughly $20\,h^{-1}$\,Mpc and above). 
We note that these distortions can be corrected in an iterative fashion (see the pioneering work of \citealt[][]{Yahiletal1991} within the linear approximation; \citealt[][]{Kitauragalleranietal2012} for the lognormal model, or  \citealt{Kitauraerdogduetal2012} including non-local tidal fields). Nevertheless, our tests performing such kinds of correction on the mock catalogue yield a very minor improvement on the scales of interest.

\begin{figure}
\centering
\includegraphics[width=10.cm]{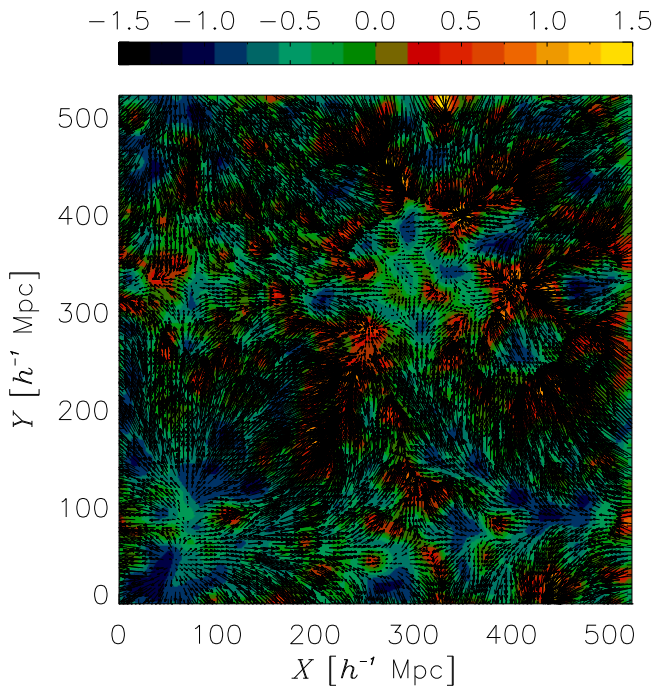}
\caption[fig:velrec]{ Reconstruction of the $x$-component of the peculiar velocity field (arrows) in a narrow slice of the grid containing the CGC over the corresponding density contrast contour 2D plot.  }
\label{fig:velrec}
\end{figure}

\begin{figure}
\centering
\includegraphics[width=9.cm]{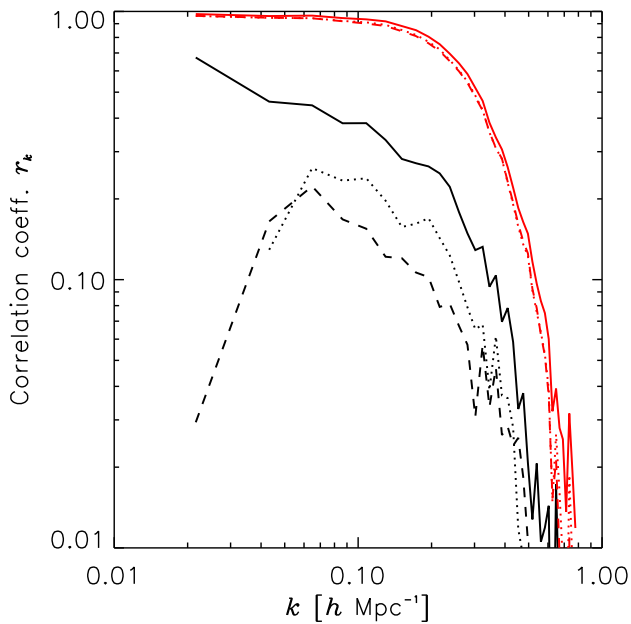}
\caption[fig:rk]{ Correlation coefficient of the recovered line-of-sight velocities with the actual ones in our \galaxy\ mock catalogue. Solid, dotted, and dashed lines refer to the LINEAR, LOG-LINEAR, and LOG-2LPT approaches, respectively. The red lines consider the ideal scenario without any sky mask or selection function, while the black ones are for the same sky mask and selection function present in the real CGC.}
\label{fig:rk}
\end{figure}

After aligning the $X$- and $Z$-axes of the 3D grid with the zero Galactic longitude and zero Galactic co-latitude axes, respectively, we place our grid at a distance vector $\vec{R}_{\rm box}= [-300, -250, 150]$\, $h^{-1}$\,Mpc from the observer. This position vector locates the corner of the 3D grid that constitutes the origin for labelling cells within the box. This choice of $\vec{R}_{\rm box}$ is motivated by a compromise between  having as many grid cells in the CGC footprint as possible, and keeping a relatively high galaxy number density. Placing the 3D grid at a larger distance would allow us to have all grid cells inside the CGC footprint, but at the expense of probing distances where the galaxy number density is low (due to the galaxy radial selection function being low as well). Our choice for ${\vec R}_{\rm box}$ results in about 150\,000 CGs being present in the box, and about 82\,\% of the grid cells falling inside the CGC footprint. 

The three approaches will provide estimates of the peculiar velocity field in each grid cell, $\vv^{\rm rec}(\vx )$. From these, it is straightforward to compute the radial component as seen by the observer, $v_{\rm los}^{\rm rec} (\vx)$, and to assign it to all galaxies falling into that grid cell.
In Fig.~\ref{fig:velrec} arrows show the LINEAR reconstruction of the $x$-component of the peculiar velocity field from the CGC, for a single $z$-slice of data. The coloured contour displays the galaxy density contrast distribution over the same spatial slice.

The methodology outlined in K12 was conducted in the absence of any sky mask or selection function. In our work we address these aspects of the real data by means of a Poissonian data augmentation approach. In a first step, all grid cells falling outside the CGC footprint are populated, via Poissonian realizations, with the average number of galaxies dictated by the CGC radial selection function computed from cells inside the footprint. In this way we fill all holes in the 3D grid. Following exactly the same procedure, we next radially augment the average number of counts in cells in such a way that the radial selection function of the resulting galaxy sample is constant. In this way we avoid radial gradients that could introduce spurious velocities along the line of sight. 

In order to test this methodology, we use the \galaxy\ mock catalogue, in two different scenarios. The first scenario is an ideal one, with no mask or selection function: we apply the three approaches on a box populated with our mock catalogue, and compare the recovered radial velocities with the real ones provided by the catalogue. When performing this comparison, we evaluate a 3D grid for the original galaxy radial velocity $\vloso$ by assigning to each cell the radial velocity of the galaxy falling nearest to the cell centre, that is, we adopt a {\it nearest particle} (NP) method. For different radial bins of the $\vk$ wavevector, we evaluate the correlation coefficient ($r_k$) computed from the cross-power spectrum between the recovered and the original radial velocity fields and their respective auto-spectra:
\begin{equation}
r_k = \frac{P^{\rm \,orig,\,rec}(k)}{\sqrt{P^{\rm \,orig,\, orig}(k)\, P^{\rm \,rec,\,rec}(k)}}.
\label{eq:rk}
\end{equation}
In this expression, $P^{X,\,Y}(k)$ stands for $\langle v(\vk)_{\rm los}^{ X} (v(\vk)_{\rm los}^{ Y})^\ast \rangle$, that is, the radially averaged power spectrum of the line of sight velocity modes. The superscripts $\{ X,\,Y\}=\{ {\rm orig, rec}\}$ stand for ``original" and ``recovered" components, respectively, and the asterisk denotes the complex conjugate operation. 

In Fig.~\ref{fig:rk} the red lines display the correlation coefficients for the ideal scenario. Solid, dotted, and dashed lines refer to the LINEAR, LOG-LINEAR, and LOG-2LPT approaches, respectively. On large scales (low $k$ modes), the three approaches provide correlation coefficients that are very close to unity, while they seem to lose information on small scales in the same way.  We note that a direct comparison to the results of K12 is not possible, since, in our case, we do not use the full dark matter particle catalogue, but only a central halo one.

After the real mask and selection function obtained from the CGC is applied to the \galaxy\ mock catalogue, then the recovery of the peculiar velocities from the three adopted approaches worsens considerably. The black lines in Fig.~\ref{fig:rk} display, in general, much lower correlation levels than the red ones. The LINEAR approach seems to be the one that retains most information on the largest scales, and it out-performs the LOG-LINEAR and the LOG-2LPT methods on all scales. We hence expect the LINEAR approach to be more sensitive to the kSZ effect than the other two methods, particularly on the largest scales.

Once the velocity inversion from the CGC density contrast has been performed, we compute the spatial correlation function between the recovered velocities and the kSZ temperature anisotropies,
\begin{equation}
w^{T,v} (r) = \langle \delta T_i \vlosr (\vx_j) \rangle_{i,j} (r),
\label{eq:Ctv}
\end{equation}
 where $\delta T_i$ is estimated as in Eq.~(\ref{eq:dT}), and the ensemble average is obtained after running through all galaxy pairs $\{i,j\}$ lying a distance $r$ away.
 
\subsection[]{Template fitting}

When studying the measurements of the kSZ pairwise momentum and kSZ momentum-$\vlosr$ correlation, we perform fits to estimates obtained from our numerical simulations. That is, we minimize the quantity
\begin{equation}
\chi^2 = \sum_{ i,j} \bigl(\hat{w}^{ X}(r_{ i}) - A^{ X}\, \tilde{w}^{ X,\,{\rm sim}}(r_{ i})\bigr)\, \tens{C}_{ i j}^{-1} \, \bigl(\hat{w}^{ X}(r_{ j}) - A^{ X}\, \tilde{w}^{ X,\,{\rm sim}}(r_{ j})\bigr),
\label{eq:chis1}
\end{equation}
where the indexes ${i,j}$ run over different radial bins, $\hat{w}^{ X}(r_{ i})$ is the measured quantity in the $i$th radial bin (with ${ X}$ either denoting kSZ pairwise momentum or the kSZ temperature-recovered velocity correlation), and $\tilde{w}^{ X\, {\rm sim}}(r_{ i})$ refers to its counterpart measured in the numerical simulation. The symbol $\tens{C}_{ i j} $ denotes the $i j$ component of the covariance matrix $\tens{C}$ that is computed from the \Planck\ maps after estimating $\hat{w}^{ X}$ for {\em null} positions where no SZ effect is expected\footnote{These null positions on the \Planck\ maps correspond to rotated or displaced positions with respect to the original location of the CGs, as will be explained in the next section.}. This minimization procedure provides formal estimates for the amplitude $A^{ X}$ and its associated errors:
\begin{eqnarray}
A^{ X} & = & \frac{ \sum_{ i,j} \hat{w}^{ X}(r_{ i}) \, C_{ i,j}^{-1} \, \tilde{w}^{ X,\,{\rm sim}}(r_{ j}) }  {\sum_{ i,j} \tilde{w}^{ X,\, {\rm sim}}(r_{ i}) \, C_{ i,j}^{-1} \, \tilde{w}^{ X,\,{\rm sim}}(r_{ j})};
\label{eq:Aanderror1}
\\
\sigma^2_{A^{ X}} & = & \frac{1}
{\sum_{ i,j} \tilde{w}^{ X,\, {\rm sim}}(r_{ i}) \, C_{ i,j}^{-1} \, \tilde{w}^{ X,\,{\rm sim}}(r_{ j})  }.
\label{eq:Aanderror2}
\end{eqnarray}

\begin{figure*}
\centering
\includegraphics[width=18.cm]{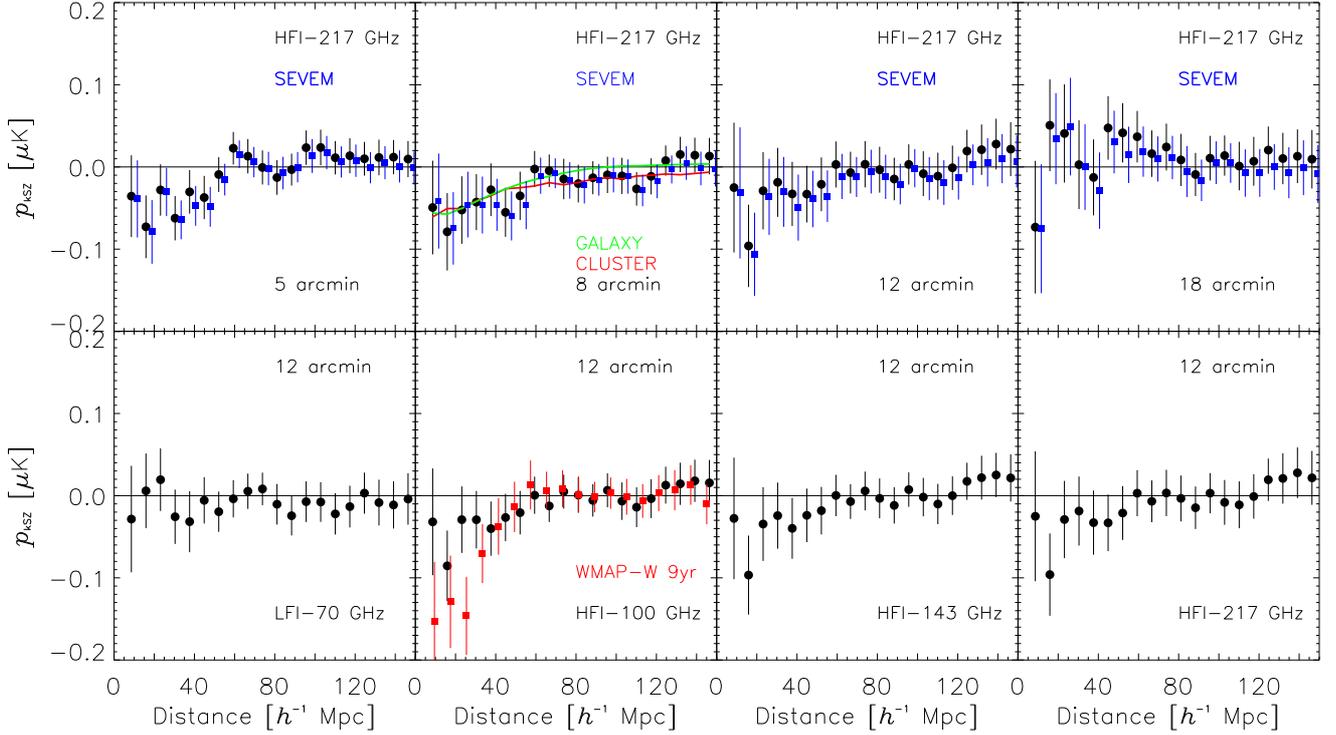}
\caption[fig:kSZ_cgc]{Computation of the kSZ pairwise momentum for the CGC sample. The top row, from left to right, displays the results for different aperture choices on the raw HFI 217\,GHz map, namely 5, 8, 12, and 18\,arcmin. The top row also shows the analysis for the foreground-cleaned \sevem\ map, displayed with blue squares. The fit to the pairwise momentum templates from the \cluster\ catalogue is also displayed by the solid red line, and the \galaxy\ catalogue by the green line. The bottom row presents the results at a fixed aperture of 12\,arcmin for different frequency maps, including \wmap-9 W-band data (red squares).   }
\label{fig:kSZ_cgc}
\end{figure*}

Most of the information is located at short and intermediate distances, where the estimated statistic $\hat{w}^{ X}(r)$ differs most from zero, as can be seen in Figs.~\ref{fig:kSZ_cgc} and \ref{fig:clusvsap}. We also test the null hypothesis, that is, we measure the $\chi^2$ statistic (defined in Eq.~\ref{eq:chis1} above) for the particular case of $A^{ X}=0$ and estimate the significance level at which such a value (denoted by $\chi^2_{\rm null}$)  is compatible with this null hypothesis. In these cases, we quote the significance as the number of $\sigma$ with which the null hypothesis is ruled out under Gaussian statistics. The array of distance bins on which the covariance matrix in Eq.~(\ref{eq:chis1}) is computed is chosen evenly in the range 0--150\,$h^{-1}$\,Mpc. However, we consider only three separate points centred upon 16, 38, and 81\,$h^{-1}$\,Mpc when computing (conservative) statistical significances. In this way we minimize correlation among radial bins to ensure that the inversion of the covariance matrix is stable. We have checked that, for the adopted set of distance bins, random variations at the level of 10\,\% of the measured $\hat{w}^{ X} (r)$ do not compromise the stability of the recovered significance estimates. In other words, we explicitly check that 10\,\% fluctuations on the measured $\hat{w}^{ X} (r)$ introduces fluctuations at a similar level in the $\chi^2$ estimates (more dramatic changes in the $\chi^2$ estimates would point to singular or quasi-singular inverse covariance matrices).

\section{Results}
\label{sec:results}

\subsection{The kSZ pairwise momentum}
\label{sec:pkSZ}

As mentioned above, the CGC was used previously in \citet{planck2012-XI} to trace the tSZ effect versus stellar mass down to halos of size about twice that of the Milky Way. In this case we use the full CGC to trace the presence of the kSZ signal in \Planck\ data, since our attempts with the most massive sub-samples of the CGC yield no kSZ signatures.
While the tSZ effect is mostly generated in collapsed structures \citep{Hernandezetal2006a} because it traces gas pressure, the kSZ effect instead is sensitive to {\em all} baryons, regardless of whether they belong to a collapsed gas cloud or not. Thus it is expected that not only will the virialized gas in halos contribute to the kSZ signal, but also the baryons surrounding those halos and moving in the same bulk flow. Rather than using a particular gas density profile, we choose not to make any assumption about the spatial distribution of gas around CGs.


We thus adopt an AP filter of varying apertures around the positions of CGC galaxies. The minimum aperture we consider is close to the resolution of \Planck\ (5\,arcmin), and we search for signals using increasing apertures of radii 5, 8, 12, and 18\,arcmin. This is equivalent to probing spheres of radius ranging from 0.5 up to 1.8\,$h^{-1}$\,Mpc (in physical units) around the catalogue objects. The result of the kSZ pairwise momentum estimation on the raw 217\,GHz HFI map is displayed in the top row of Fig.~\ref{fig:kSZ_cgc} for the four apertures under consideration. The recovered momenta for the raw frequency maps (displayed by black circles) provide some evidence for kSZ signal for apertures smaller than 12\,arcmin: below a distance of 60\,$h^{-1}$\,Mpc all points are systematically below zero, some beyond the $2\,\sigma$ level. Although the $\chi^2_{\rm null}$ test does not yield significant values (at $0.4\,\sigma, 0.3\,\sigma,0.7\,\sigma$ and $-1.2\,\sigma$ for 5, 8, 12, and 18\,arcmin apertures, respectively), the fits to the \galaxy\ peculiar momentum template (displayed by the green solid line in the second-from-the-left panel in the top row) yield S/N $=A^{p_{\rm kSZ}}/\sigma^{p_{\rm kSZ}}=1.7, 1.4, 1.9$ and 0.1 for 5, 8, 12, and 18\,arcmin apertures, respectively. We obtain similar levels for the peculiar momentum template obtained from the \cluster\ simulation (red line in second-from-the-left panel in top row). When using the \sevem\ clean map, these significances go beyond the $2\,\sigma$ level: S/N $=A^{p_{\rm kSZ}}/\sigma^{p_{\rm kSZ}}=2.5, 1.8, 2.2$ and 0.0 for 5, 8, 12, and 18\,arcmin apertures for the \galaxy-derived template, respectively. If we repeat the analysis at 12\,arcmin aperture for the \smica\, \nilc\ and \commander\ maps, we obtain S/N$=A^{p_{\rm kSZ}}/\sigma^{p_{\rm kSZ}}=2.1, 2.2,$ and 2.1, respectively.

\begin{figure}
\centering
\includegraphics[width=9.cm]{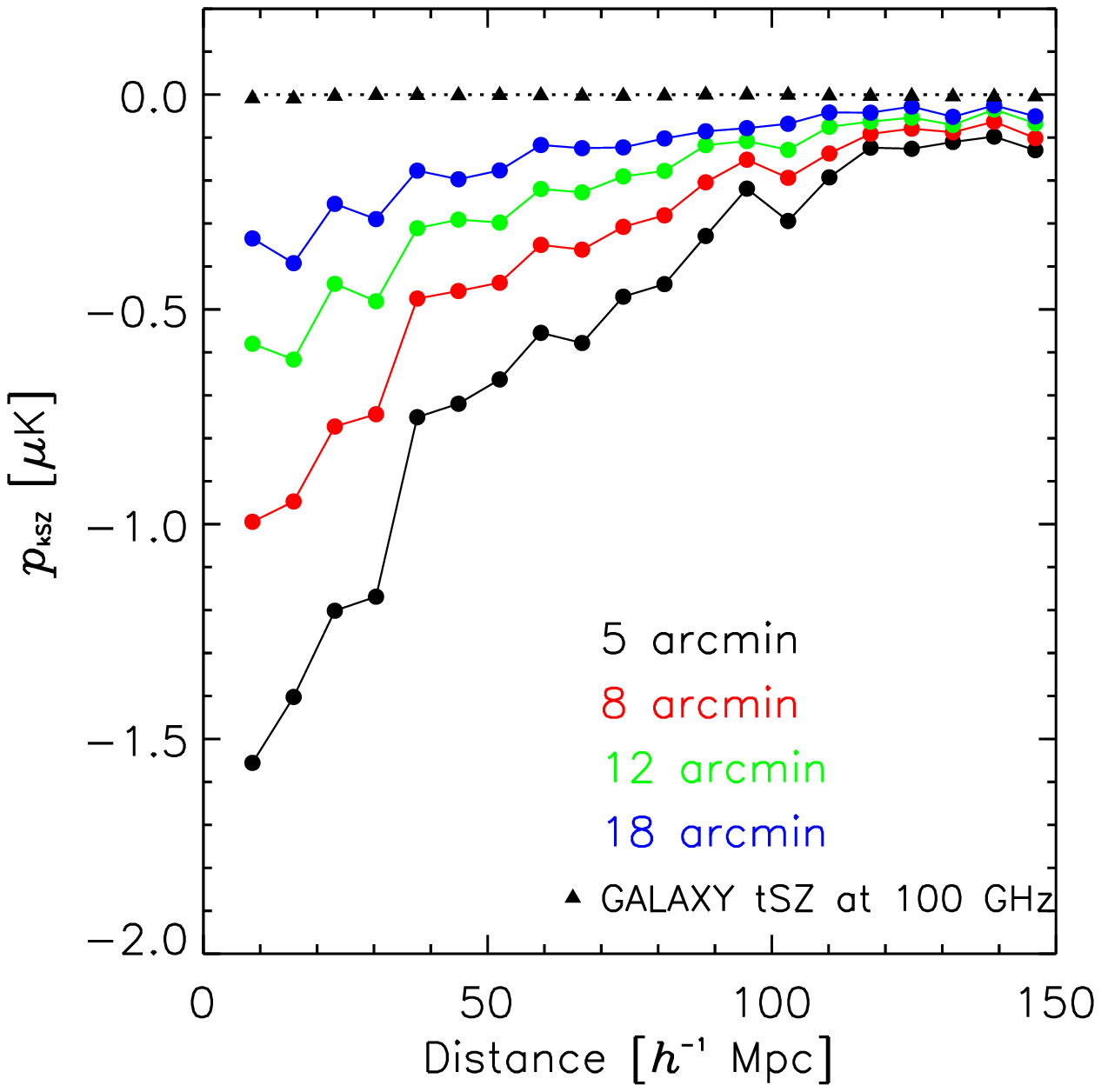}
\caption[fig:clusvsap]{  {\it (Filled circles):} Measured kSZ pairwise momentum for the kSZ map derived from the \cluster\ simulation after considering different radial apertures on a subset of clusters in the mass range (1--2)\,$\times 10^{14}$\, $h^{-1}$\,M$_{\odot}$. {\it (Black triangles):} Pairwise momentum computed from the \galaxy\ mock catalogue (filled triangles) after assigning these galaxies a tSZ amplitude following a mass scaling inspired by the tSZ measurements of the CGC given in \citet{planck2012-XI}.}
\label{fig:clusvsap}
\end{figure}

We note at this point that the behaviour of the kSZ evidence for the CGs as displayed in Fig.~\ref{fig:kSZ_cgc}  is significantly different to what is found in Fig.~\ref{fig:clusvsap} for the \cluster\ simulation with halos in the mass range $(1$--$2)\,\times\, 10^{14}\,h^{-1}$\,M$_{\odot}$. In this simulation, most of the kSZ signal is coming from the halos themselves, and thus increasing the aperture to values larger than the virial radius of the clusters results in a dilution of the kSZ pairwise momentum amplitude. For real data, we find that the amplitude of the signal does not show significant changes when increasing the aperture from 5\,arcmin up to 12\,arcmin.
It is worth adding now that the black filled triangles in Fig.~\ref{fig:clusvsap} represent the peculiar momentum measured on the \galaxy\ mock catalogue after assigning to these galaxies a tSZ temperature fluctuation following the tSZ versus mass scaling measured in the CGC \citep{planck2012-XI}. This tSZ amplitude assumes an observing frequency of $\nu_{\rm obs} = 100$\,GHz, and hence provides a conservative estimate of the tSZ contamination, which turns out to be at the level of $-3\times 10^{-3}$\,$\mu$K, i.e., about a factor of 30 below the measured amplitude on the real CG sample.

We next investigate the spectral stability of the signal for 12\,arcmin apertures. This choice is motivated by the fact that the LFI 70\,GHz and the HFI 100 and 143\,GHz channels have lower angular resolution, comparable or bigger than 8\,arcmin, and this may compromise the comparison with the 217\,GHz channel. This analysis is shown in the bottom row of Fig.~\ref{fig:kSZ_cgc}. The signal obtained for the raw 217\,GHz channel is found to be remarkably similar to what we obtain at 143, 100\,GHz in HFI: the fits to the kSZ pairwise momentum template from the \galaxy\ simulation gives S/N $=A^{\rm p_{kSZ}}/\sigma^{\rm p_{kSZ}}=2.2$ and 2.1, respectively. The LFI 70\,GHz channel, on the other hand, seems to show a much flatter pattern, and this could be due to a larger impact of instrumental noise: the tSZ should also be present at 100 and 143\,GHz, and, as expected, gives rise to negligible changes. The effective full width half maximum (FWHM) for the 70\,GHz channel is close to 13\,arcmin, and this may also contribute to the difference with respect to the other HFI channels.\\

In the bottom row, second panel of Fig.~\ref{fig:kSZ_cgc}, the red squares provide the measurement obtained from the cleaned W-band map of \wmap-9. Surprisingly, the level of anti-correlation for distances below 30\,$h^{-1}$\,Mpc appears higher for \wmap\ data than for the \Planck\ channels: the first three radial bins lie at the $2.3$--$3.3\,\sigma$ level, and a fit to the \galaxy\ template of the pairwise momentum yields S/N $=A^{\rm p_{kSZ}}/\sigma^{\rm p_{kSZ}}=3.3$ as well. 
The angular resolution of the W-band in \wmap\ is close to that of the LFI 70\,GHz channel, and the non-relativistic tSZ changes by less than 13\,\% between those channels. If any other frequency-dependent contaminants (which should be absent in the \wmap\ W-band) are responsible for this offset, then they should also introduce more changes at the other frequencies. Thus the reason for the deeper anti-correlation pattern found in the W-band of \wmap-9 remains unclear. Overall, we conclude that the \Planck results for the pairwise peculiar momentum are compatible with a kSZ signal, at a level ranging between 2 and $2.5\,\sigma$. The output from \wmap-9 W band is also compatible with a kSZ signal, with a statistical significance of $3.3\,\sigma$. 

\begin{figure}
\centering
\includegraphics[width=9.cm]{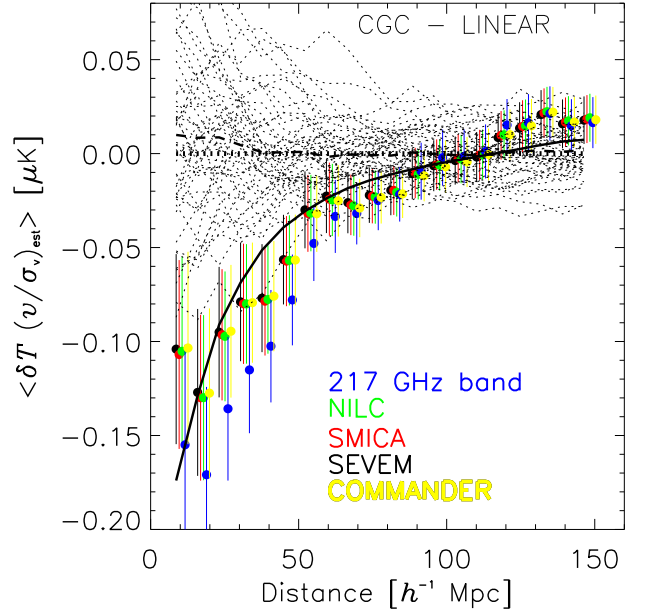}
\caption[fig:dTdv]{Measured cross-correlation function between the kSZ temperature estimates and the recovered radial peculiar velocities, $w^{{\rm T},\vlosr} (r)$, according to the LINEAR approach. The recovered velocities are divided with their rms dispersion $\sigma_{\rm v}=310\,$km\,s$^{-1}$, so kSZ temperature estimates are correlated to a quantity of variance unity.
This plot corresponds to an aperture of 8\,arcmin. Filled coloured circles correspond to  $w^{{\rm T},\vlosr} (r)$ estimates from different CMB maps (\sevem, \smica, \nilc, \commander, and the HFI 217\,GHz map). The dotted lines display the null estimates obtained after computing kSZ temperature estimates for rotated positions on the \sevem\ map, and the thick  dashed line displays the average of the dotted lines. Error bars are computed from these null estimates of the correlation function. The solid line provides the best fit of the \sevem\ data to the theoretical prediction for $w^{{\rm T},\vlosr} (r)$ obtained from the \galaxy\ mock catalogue. These predictions are obtained using only a relatively small number of mock halos, and hence their uncertainty must be considered when comparing to the data.}
\label{fig:dTdv}
\end{figure}

\subsection{Cross-correlation analysis with estimated peculiar velocities}
\label{sec:xcorrvelsTs}

\begin{figure}
\centering
\includegraphics[width=9.cm]{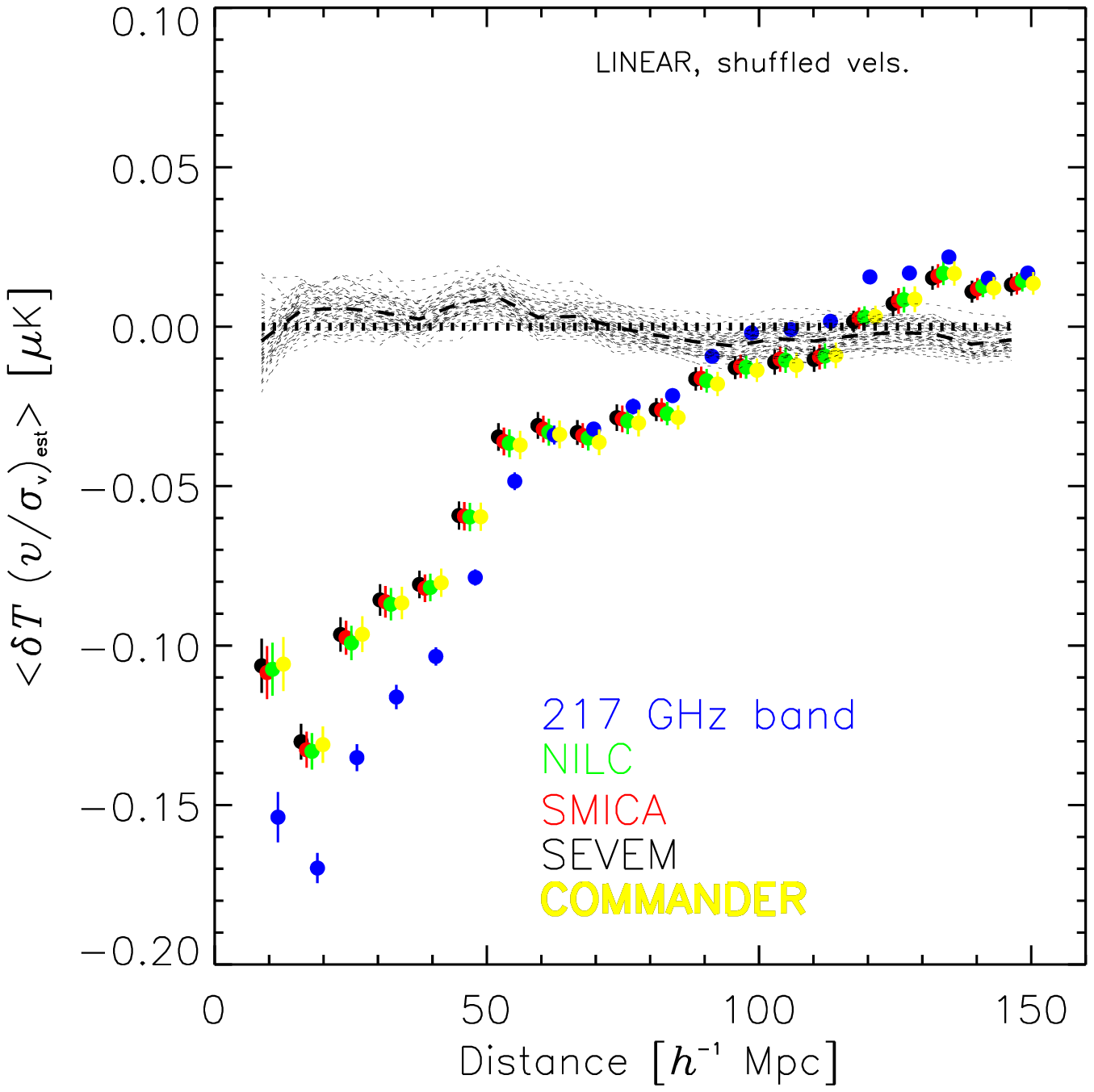}
\caption[fig:shuff]{Same as in Fig.~\ref{fig:dTdv}, but with the null correlation functions (displayed by dotted lines) being estimated after assigning the recovered linear velocity estimate of a given CG to any other CG in the sample (that is, by $\vlosr$ ``shuffling"). Only the LINEAR approach is displayed in this plot.
 }
\label{fig:shuff}
\end{figure}

\begin{figure}
\centering
\includegraphics[width=9.cm]{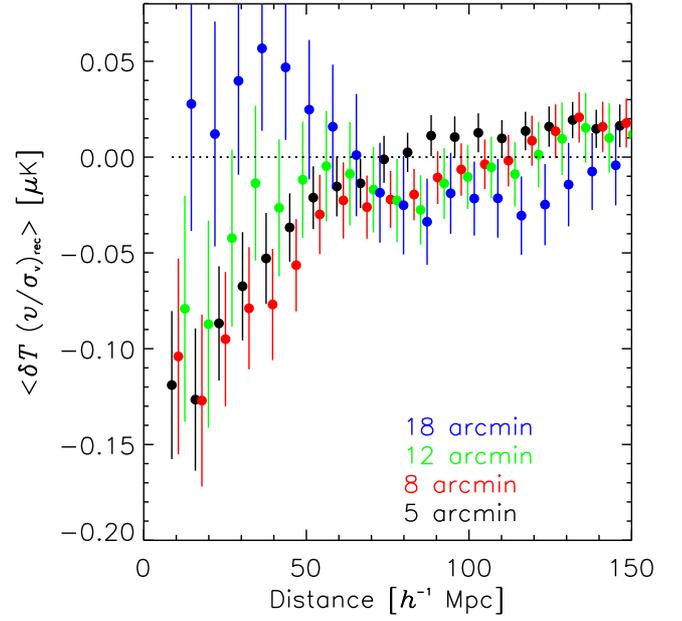}
\caption[fig:wvsAPrad]{ Dependence on the aperture radius of the $w^{{\rm T},\vlosr} (r)$ correlation function obtained in the LINEAR approach for the \sevem\ map. Error bars are estimated from the rms of the null correlation functions computed from rotated kSZ temperature estimates. }
\label{fig:wvsAPrad}
\end{figure}
\begin{figure}
\centering
\includegraphics[width=9.cm]{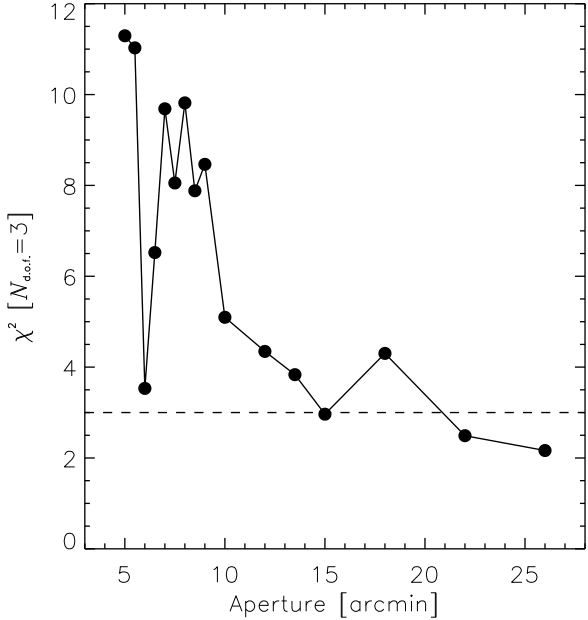}
\caption[fig:chisqvsap]{ Evaluations of the $\chi^2$ statistic of the $w^{{\rm T},\vlosr} (r)$ correlation function (with respect to the null hypothesis) for different angular apertures. Only results for the foreground-cleaned \sevem\ map are shown. There is evidence for kSZ signal in a wide range of apertures above the FWHM, with a local maximum close to 8\,arcmin. This roughly corresponds to a radius of 0.8\,$h^{-1}$\,Mpc from the CG positions at the median redshift of the sample. }
\label{fig:chisqvsap}
\end{figure}

\begin{figure}
\centering
\includegraphics[width=9.cm]{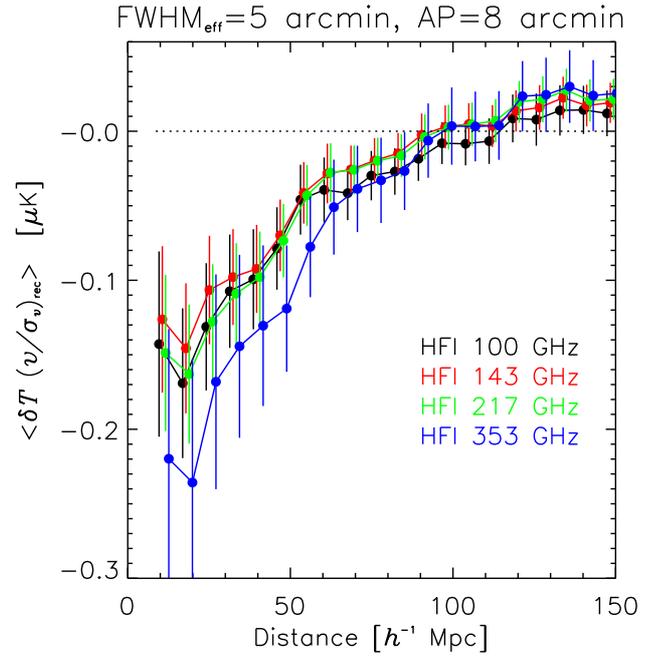}
\caption[fig:wTv]{Variation of the recovered $w^{{\rm T},\vlosr} (r)$ correlation function for the four lowest frequency raw HFI maps after an effective convolution by a Gaussian beam of FWHM $=5$\,arcmin. Error bars here are computed in the same way as in Figs.~\ref{fig:dTdv} and \ref{fig:wvsAPrad}. }
\label{fig:wTv}
\end{figure}

\begin{figure}
\centering
\includegraphics[width=9.cm]{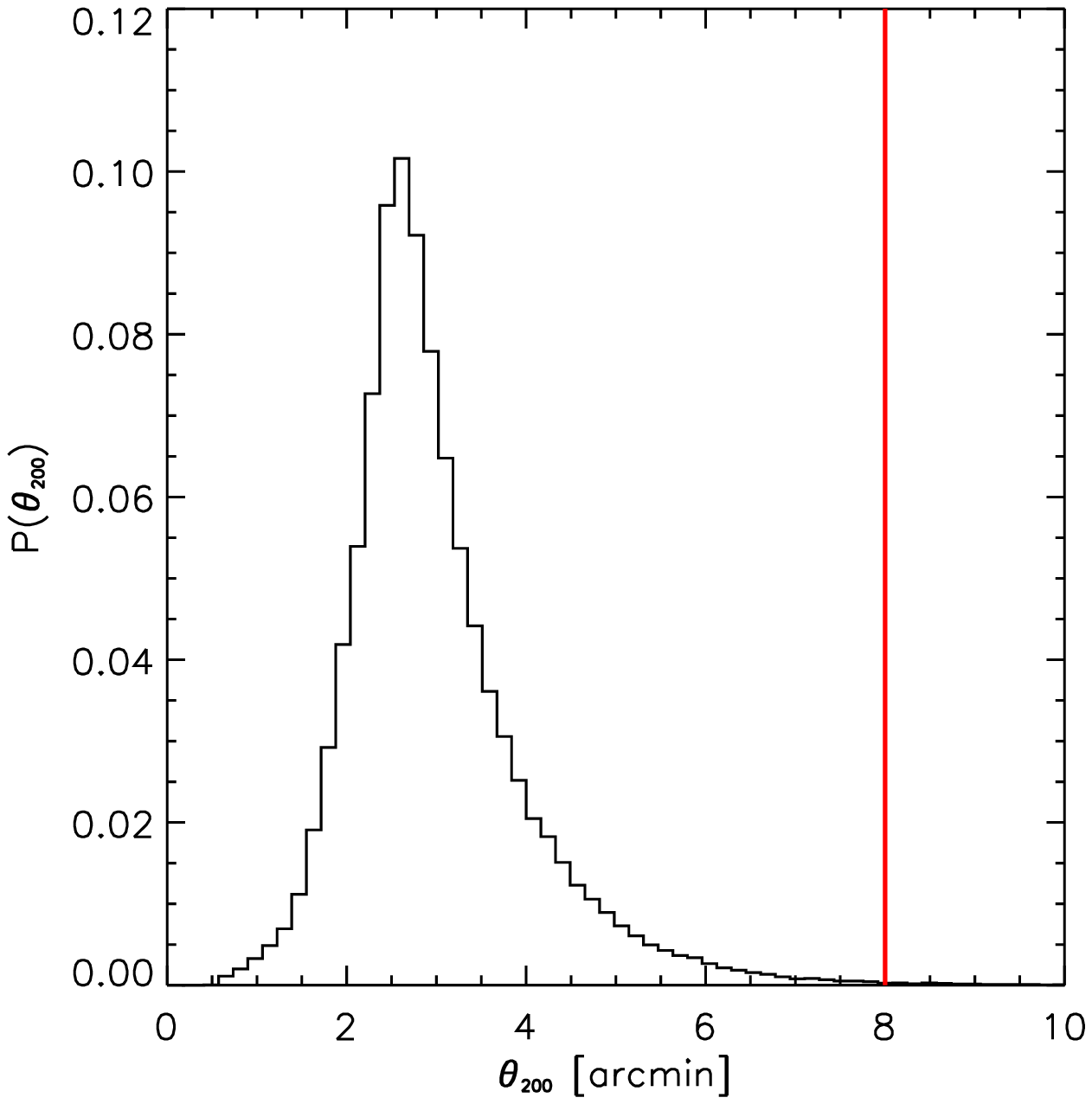}
\caption[fig:hist_Theta200]{Histogram of the angle subtended by $R_{200}$ of the CGC members.}
\label{fig:hist_Theta200}
\end{figure}


After reconstructing the CGC density field into a peculiar velocity field in a 3D grid, as explained in Sect.~\ref{sec:invertrho}, we compute the cross-correlation function $w^{{\rm T},\vlosr} (r)$ for the three inversion approaches using the \Planck\ HFI 217\,GHz band, and the \sevem\ , \smica\ , \nilc\ , and \commander\ maps (see Fig.~\ref{fig:dTdv}). Here we are using kSZ temperature estimates obtained with an 8\,arcmin aperture. The three inversion approaches (namely, LINEAR, LOG-LINEAR, and LOG-2LPT) may suffer from different biases that impact the amplitude of the recovered velocities. Thus, in order to avoid issues with the normalization of the recovered velocities and to ease the comparison between different approaches, we normalize the recovered velocities $\vlosr$-s by their rms before computing the cross-correlation function. In any case we must keep in mind the different responses of the three approaches on different $k$-scales (see Fig.~\ref{fig:rk}), thus giving rise to different correlation structures. For the sake of clarity, in this section we show results for the LINEAR method only, while in the Appendix we discuss the results for the other two reconstruction methods.

In Fig.~\ref{fig:dTdv}, filled coloured circles display the results obtained after using the $\delta T$ kSZ temperature estimates obtained at the real positions of the CGs on the CMB maps. The null tests were obtained by computing the cross-correlation of the {\em normalized} recovered velocities ($\vlosr$) with temperature estimates  ($\delta T$) obtained on 50 {\em rotated}\footnote{For each rotation/displacement we use a step of three times the aperture radius adopted.} positions on the CMB maps. In those cases, the $\delta T$s were computed for positions obtained after rotating the real CG angular positions in Galactic longitude. The results from each of these null rotations are displayed by dotted lines in Fig.~\ref{fig:dTdv}, and their average is given by the thick, dashed line (which lies close to zero at all radii). 
Both the error bars and the covariance matrix of the correlation function were obtained from these null realizations. 

Results at zero-lag rotation (i.e., the real sky) lie far from the distribution of the null rotations. There are several individual distance bins lying more than $3\,\sigma$ (up to $3.8\,\sigma$ on the $16\,h^{-1}$\,Mpc distance bin for the raw 217\,GHz frequency map), and the $\chi^2_{\rm null}$ tests rule out the null hypothesis typically at the level of $2.1$--$2.6\,\sigma$ for the clean maps, and at $3.2\,\sigma$ for the raw 217\,GHz frequency map. Likewise, when fitting the observed correlation function to the $w^{{\rm T},\vlosr} (r)$ correlation function obtained from the \galaxy\ simulation, we obtain values of S/N $=A^{w^{T,v}}/\sigma^{w^{T,v} }=3.0$--$3.2$ for the clean maps, while for the HFI 217\,GHz raw map we find S/N $=A^{w^{T,v}}/\sigma^{w^{T,v} }=3.8$. We explain this apparent mismatch below.
A clear, large-scale correlation pattern, extending up to about $80\,$ $h^{-1}$\,Mpc, is found in the data.
A complementary systematic test can be conducted by computing the cross-correlation function of the kSZ temperature fluctuations ($\delta T$) with {\em shuffled} estimates of the recovered line-of-sight peculiar velocities, i.e., to each CG we assign a $\vlosr$ estimate corresponding to a different, randomly selected CG. The result of performing this test for the LINEAR approach is displayed in Fig.~\ref{fig:shuff}, and shows that the correlation found between the $\delta T$s and the recovered velocities clearly vanishes for all shuffled configurations. By shuffling the recovered velocities we are destroying their coherent, large-scale pattern, which couples with large angle CMB residuals, and generates most of the uncertainty in the measured cross-correlation. This explains the smaller error bars in Fig.~\ref{fig:shuff} when compared to Fig.~\ref{fig:dTdv}.

We further study the dependence of the measured $w^{{\rm T},\vlosr} (r)$ correlation function on the aperture radius and show the results in Fig.~\ref{fig:wvsAPrad}. We again restrict ourselves to the LINEAR approach and the \sevem\ foreground-cleaned map, and error bars are computed from the null rotations (rather than from shuffling estimates of $\vlosr$). We display results for four apertures ranging from 5 to 18\,arcmin: the lower end of this range is given by the angular resolution of the map, while for the higher end the kSZ signal is found to vanish. We find that, as for the kSZ peculiar momentum, the kSZ amplitude at 8\,arcmin aperture is very close to the one found at 5\,arcmin, and it is still significant at 12\,arcmin aperture.
We then vary the aperture from 5 to 26\,arcmin and calculate the corresponding $\chi^{2}$ values, as shown in Fig.~\ref{fig:chisqvsap}. These results are for the \sevem\ foreground-cleaned map, and provide another view of the angular extent of the signal. For the kSZ peculiar momentum, we have shown in Fig.~\ref{fig:kSZ_cgc} that there is kSZ evidence for apertures as large as 12\,arcmin, in good agreement with what we find now for the  $w^{{\rm T},\vlosr} (r)$ correlation function. While this statistic seems to have higher significance than the pairwise momentum, for both statistics we find consistently that most of its signal is again coming from gas not locked in the central regions of halos, but in the intergalactic medium surrounding the CG host halos.

We also test the consistency of our results with respect to frequency on raw \Planck\ raw maps. For that we use the HFI channels ranging from 100\,GHz up to 353\,GHz and now we fix the aperture at 8\,arcmin. Since the 100 and 143\,GHz frequency maps have angular resolution comparable or worse than 8\,arcmin, we choose to deconvolve all HFI maps under consideration by their respective (approximate) Gaussian beams, and convolve them again with a Gaussian beam of FWHM$=5$\,arcmin. While this approach may challenge the noise levels for the HFI maps with coarser beams, we find that this is balanced by the large number of CGs on which we compute the $w^{{\rm T},\vlosr} (r)$ correlation function. The results of this analysis are given in Fig.~\ref{fig:wTv}, and show how the $w^{{\rm T},\vlosr} (r)$ correlation functions from 100\,GHz up to 217\,GHz agree closely with each other, with no significance dependence on frequency (as is expected for the kSZ effect). However, the result for the 353\,GHz channel is pointing to a significantly higher amplitude of the correlation function, even if the error bars associated with this map are 
typically 50--70\,\% larger than for the 100\,GHz map. At the map level, the main difference between the 353\,GHz and lower frequency channels is the significantly larger amount of dust and/or CIB emission in the former map. We have checked that, throughout the rotated configurations, the HFI 353\,GHz map has on average no correlation with the estimated radial velocities of the CGs. Therefore, the excess found in the measured amplitude of the $w^{{\rm T},\vlosr} (r)$ correlation function must be due to fortuitous alignment between the estimated radial velocities of the CGs and the dust emission at the position of the CGs. This would also explain the higher amplitude of the $w^{{\rm T},\vlosr} (r)$ correlation function found for the raw HFI 217\,GHz map with respect to all other clean maps, as shown in Fig.~\ref{fig:dTdv}: the raw HFI 217\,GHz map should still contain some non-negligible dust contamination when compared to the foreground-cleaned maps \sevem, \smica, \nilc, and \commander. 

Another potential issue is related to the fact that, as shown in \citet{planck2012-XI}, about 12\,\% of the CGC members with stellar masses above $10^{10}$\,M$_{\odot}$ are actually not central galaxies, but show some offset with respect to the centres of the host halos. This offset should cause a low bias in the kSZ amplitude estimation. We have simulated the impact of this offset by assuming that 12\,\% of the CGs have uniform probability of lying within a (projected) distance range of 0 to 1\,Mpc from the halo centre. After adopting a NFW profile for the gas distribution, we have found that the LOS-projected kSZ signal should be low biased a 10, 8, and 6\,\% for $\theta_{\rm AP}=5,10$ and 15\,arcmin, respectively. Given the limited precision of our measurements, these biases are relatively small and will be ignored hereafter. 

To compare our detection with simulations, the solid line in Fig.~\ref{fig:dTdv} shows the best fit to the prediction inferred from the \galaxy\ mock catalogue. This prediction is obtained by applying the LINEAR velocity recovery algorithm on our \galaxy\ catalogue after imposing the sky mask and the selection function present in the real CGC. Three pseudo-independent estimates of the correlation function of the recovered LOS velocity $v_{\rm rec}$ and the real LOS velocity $v_{\rm los}$  ($\langle v_{\rm los} v_{\rm rec}\rangle (r) \equiv w^{v_{\rm rec},v}(r)$) are obtained after rotating the 3D grid hosting the mock \galaxy\ CGC, so that the side facing the observer is different in each case. The solid line corresponds to the average of these three estimates of $w^{v_{\rm rec},v}(r)$, and the ratio to the observed correlation function can be interpreted as an ``effective" optical depth to Thomson scattering: $w^{{\rm T},\vlosr} (r) = -\tau_{\rm
T} \,w^{v_{\rm rec},v}(r)$. We obtain $\tau_{\rm T}=(1.39\pm 0.46)\times 10^{-4}$ (i.e., at the $3\,\sigma$ level) for the \sevem\ map, with very similar values for all other foreground-cleaned maps. We defer the physical interpretation of the kSZ measurements of this paper to an external publication, \citep[]{chm_prl15}.

\section{Discussion and Conclusion}
\label{sec:conclusions}

The roughly $2.2\,\sigma$ detection (varying slightly for the different maps) of the pairwise momentum indicates that the baryonic gas is comoving with the underlying matter flows, even though it may lie outside the virial radius of the halos. The aperture of 8\,arcmin on the CGs (placed at a median redshift of $\bar{z}=0.12$) corresponds to a physical radius of around 1\,Mpc. As we show next, this is considerably higher than the typical virial radius of the CG host halos. Following the same approach as in \citet{planck2012-XI}, we compute the $R_{200}$ radius containing an average matter density equal to 200 times the critical density at the halo's redshift. In Fig.~\ref{fig:hist_Theta200} we display the histogram of the angle subtended by the $R_{200}$ values of the 150\,000 CGs placed in the 3D grid that we use for the velocity recovery. The red vertical solid line indicates the 8\,arcmin aperture, well above the typical angular size of the CGC sources. 

The behaviour displayed by the measured kSZ peculiar momentum in the top row panels of Fig.~\ref{fig:kSZ_cgc} differs significantly from the pattern found in Fig.~\ref{fig:clusvsap}. The fact that the measured kSZ pairwise momentum shows a roughly constant amplitude out to an aperture of 12\,arcmin, well above the CG virial size, signals the presence of unbound gas that is contributing to the measurement. The opposite situation is seen in Fig.~\ref{fig:clusvsap}: in this scenario most of the signal comes from the halos rather than from a surrounding gas cloud. This plot displays the kSZ peculiar momentum from a subset of our \cluster\ catalogue, with sources in the range (1--2)\,$\times 10^{14}$\, $h^{-1}$\,M$_\odot$, after considering different aperture radii. As long as the halo remains unresolved (as is the case for these simulated clusters), then as we increase the aperture size the kSZ signal coming from the halo becomes more diluted and hence the amplitude decreases, contrary to what is found in Fig.~\ref{fig:kSZ_cgc}.

We find a similar situation in the kSZ temperature-peculiar velocity cross-correlation. By cross-correlating the reconstructed peculiar velocity field in a 3D box with the kSZ temperature anisotropies, we find a 3.0\,$\sigma$ detection between the two fields for the \sevem\ map, at an aperture of  8\, arcmin. This again, corresponds to gas clouds with radius roughly $1\,$Mpc, about twice the mean $R_{200}$ radius (we find that $\langle R_{200}\rangle_{\rm CGC}\simeq 0.4\,$Mpc). Since the peculiar velocity is directly related to the underlying matter distribution, our result suggests that gas inside and outside CG host halos are comoving with the matter flows.

One way of quantifying the amplitude of our signal is to ask by what
factor we need to scale the model-based $w^{v_{\rm rec},v}(r)$ in order to match our
measured $w^{{\rm T},\vlosr} (r)$.  Interpreting this scaling as an ``effective"
optical depth to Thomson scattering we find $\tau_{\rm T}=(1.4\pm 0.5)\times 10^{-4}$,
which is a factor of 3 larger than that expected for the gas in typical CG host halos
alone. This provides another piece of evidence that the kSZ signal found
in \Planck\ data is generated by gas beyond the virialized regions around the CGs, as
opposed to the tSZ effect, which is mostly generated inside collapsed structures
\citep{Hernandezetal2006a,vanwaerbekeetal14,maetal2014} .

\begin{acknowledgements}

The Planck Collaboration acknowledges the support of: ESA; CNES, and
CNRS/INSU-IN2P3-INP (France); ASI, CNR, and INAF (Italy); NASA and DoE
(USA); STFC and UKSA (UK); CSIC, MINECO, JA and RES (Spain); Tekes, AoF,
and CSC (Finland); DLR and MPG (Germany); CSA (Canada); DTU Space
(Denmark); SER/SSO (Switzerland); RCN (Norway); SFI (Ireland);
FCT/MCTES (Portugal); ERC and PRACE (EU). A description of the Planck
Collaboration and a list of its members, indicating which technical
or scientific activities they have been involved in, can be found at
\url{http://www.cosmos.esa.int/web/planck/}. This research was supported by ERC Starting Grant (no.~307209), by the Marie Curie Career Integration Grant CIG 294183 and the by the Spanish Ministerio de Econom\'\i a e Innovaci\'on project AYA2012-30789.

\end{acknowledgements}

\bibliographystyle{aa}
\bibliography{pip105,Planck_bib}

\appendix
\section{The three velocity reconstruction methods}
\label{sec:AppA}

\begin{figure*}
\centering
\includegraphics[width=18.cm]{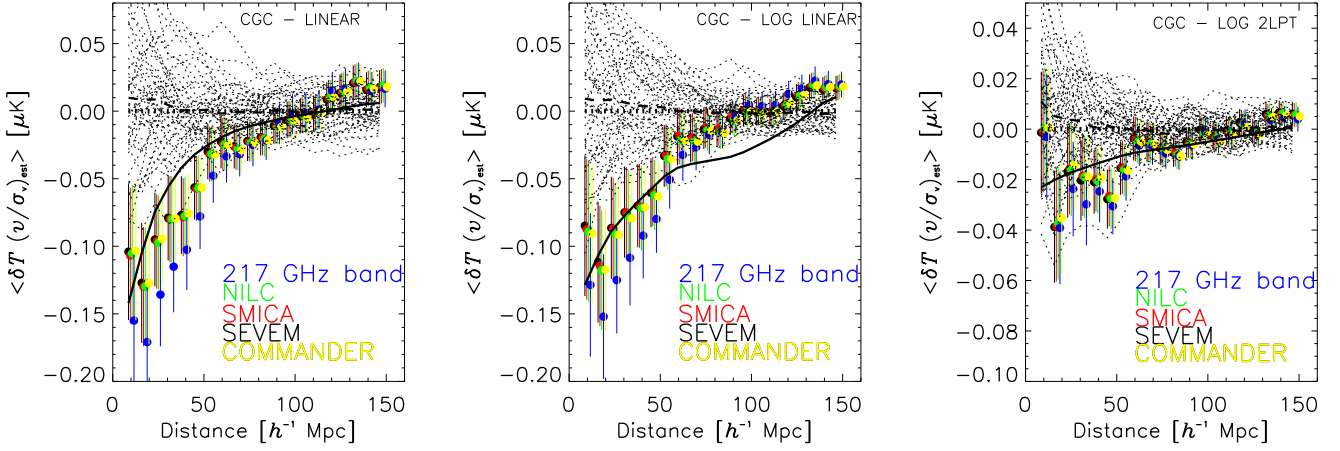}
\caption[fig:dTdv_app]{Measured cross-correlation function between the kSZ temperature estimates and the recovered radial peculiar velocities, $w^{{\rm T},\vlosr} (r)$, according to the three adopted approaches, LINEAR (left panel), LOG-LINEAR (middle panel), and LOG-2LPT (right panel). The velocities are normalized with their rms dispersion $\sigma_{\rm v}=230$, $310$, and $200$\,km\,s$^{-1}$ from left to right, respectively. We use an aperture of 8\,arcmin when estimating kSZ temperature fluctuations. Filled coloured circles correspond to  $w^{{\rm T},\vlosr} (r)$ estimates from different CMB maps (\sevem, \smica, \nilc, \commander, and the HFI 217\,GHz map). The dotted lines display the null estimates obtained after computing kSZ temperature estimates for rotated positions on the \sevem\ map, and the thick  dashed line displays the average of the dotted lines. Error bars are computed from these null estimates of the correlation function. The solid line provides the best fit to the data of the theoretical prediction for $w^{{\rm T},\vlosr} (r)$ obtained from the \galaxy\ mock catalogue. These predictions are obtained using only a relatively small number of mock halos, and hence their uncertainty must be considered when comparing to data.}
\label{fig:dTdv_app}
\end{figure*}

In this appendix we briefly compare the velocity recovery by means of the three approaches explained in this work, namely the LINEAR, the LOG-LINEAR, and the LOG-2LPT approaches. First of all, as is shown in Fig.~\ref{fig:rk}, the LINEAR method outperforms our implementation of the other two methods in the presence of a mask and selection function (even if all three provide virtually identical correlation coefficients in the absence of a mask and selection functions). This is shown clearly in Fig.~\ref{fig:dTdv_app}.

We apply our algorithms to invert density into velocities after using the three different sides of the cubic grid where we place our \galaxy\ mock catalogue. For each box side we estimate a recovered velocity field and compute its correlation to the real, underlying velocity field of the mock galaxies, $w^{v_{\rm rec},v}(r)$.
Three different estimations are too few to an average correlation function, but  we still use this average function (out of these three box-side estimates) when conducting the fit to real data. Thus the solid lines in Fig.~\ref{fig:dTdv_app} show the fits to the result of averaging the correlation functions $w^{{\rm T},v}$ through the three box orientations. We must keep in mind that we expect a high level of uncertainty in these predictions. 

These average $w^{{\rm T},v}(r)$ functions show a smooth behaviour for the LINEAR and LOG-2LPT implementations, although the LOG-LINEAR case displays a slight flattening on 50--80\,$h^{-1}$\,Mpc scales, which is absent in the other two cases and is not well fit by the data. We expect correlation functions obtained from the \galaxy\ mock catalogue to be noisier than the real ones obtained from the CGC, since typically only about 80\,000 mock galaxies survive the mask and the selection function, as opposed to the 150\,000 surviving galaxies in the real CG sample. The fits of the real data to the solid lines in Fig.~\ref{fig:dTdv_app} yield S/N $\equiv A^{ w^{v,{\rm T}}}/\sigma_{A^{ w_{v,{\rm T}}}} = $ 3.0, 2.0, and 2.1 for the LINEAR, LOG-LINEAR, and LOG-2LPT approaches, respectively.

\raggedright
\end{document}

%% file: Planck.tex
\def\setsymbol#1#2{\expandafter\def\csname #1\endcsname{#2}}
\def\getsymbol#1{\csname #1\endcsname}

\def\Planck{\textit{Planck}}

\def\all2013resultspapers{\nocite{planck2013-p01, planck2013-p02, planck2013-p02a, planck2013-p02d, planck2013-p02b, planck2013-p03, planck2013-p03c, planck2013-p03f, planck2013-p03d, planck2013-p03e, planck2013-p01a, planck2013-p06, planck2013-p03a, planck2013-pip88, planck2013-p08, planck2013-p11, planck2013-p12, planck2013-p13, planck2013-p14, planck2013-p15, planck2013-p05b, planck2013-p17, planck2013-p09, planck2013-p09a, planck2013-p20, planck2013-p19, planck2013-pipaberration, planck2013-p05, planck2013-p05a, planck2013-pip56, planck2013-p06b}}

\newbox\tablebox    \newdimen\tablewidth
\def\leaderfil{\leaders\hbox to 5pt{\hss.\hss}\hfil}
\def\tablenote#1 #2\par{\begingroup \parindent=0.8em
    \abovedisplayshortskip=0pt\belowdisplayshortskip=0pt
    \noindent
    $$\hss\vbox{\hsize\tablewidth \hangindent=\parindent \hangafter=1 \noindent
    \hbox to \parindent{$^#1$\hss}\strut#2\strut\par}\hss$$
    \endgroup}

\def\L2{\ifmmode L_2\else $L_2$\fi}

\def\DeltaT{\ifmmode \Delta T\else $\Delta T$\fi}
\def\deltat{\ifmmode \Delta t\else $\Delta t$\fi}
\def\fknee{\ifmmode f_{\rm knee}\else $f_{\rm knee}$\fi}
\def\Fmax{\ifmmode F_{\rm max}\else $F_{\rm max}$\fi}
\def\solar{\ifmmode{\rm M}_{\mathord\odot}\else${\rm M}_{\mathord\odot}$\fi}
\def\Msolar{\ifmmode{\rm M}_{\mathord\odot}\else${\rm M}_{\mathord\odot}$\fi}
\def\Lsolar{\ifmmode{\rm L}_{\mathord\odot}\else${\rm L}_{\mathord\odot}$\fi}

\def\inv{\ifmmode^{-1}\else$^{-1}$\fi}
\def\mo{\ifmmode^{-1}\else$^{-1}$\fi}
\def\sup#1{\ifmmode ^{\rm #1}\else $^{\rm #1}$\fi}
\def\expo#1{\ifmmode \times 10^{#1}\else $\times 10^{#1}$\fi}
\def\,{\thinspace}
\def\lsim{\mathrel{\raise .4ex\hbox{\rlap{$<$}\lower 1.2ex\hbox{$\sim$}}}}
\def\gsim{\mathrel{\raise .4ex\hbox{\rlap{$>$}\lower 1.2ex\hbox{$\sim$}}}}

\def\simprop{\mathrel{\raise .4ex\hbox{\rlap{$\propto$}\lower 1.2ex\hbox{$\sim$}}}}
\def\deg{\ifmmode^\circ\else$^\circ$\fi}
\def\pdeg{\ifmmode $\setbox0=\hbox{$^{\circ}$}\rlap{\hskip.11\wd0 .}$^{\circ}
          \else \setbox0=\hbox{$^{\circ}$}\rlap{\hskip.11\wd0 .}$^{\circ}$\fi}
\def\arcs{\ifmmode {^{\scriptstyle\prime\prime}}
          \else $^{\scriptstyle\prime\prime}$\fi}
\def\arcm{\ifmmode {^{\scriptstyle\prime}}
          \else $^{\scriptstyle\prime}$\fi}
\newdimen\sa  \newdimen\sb
\def\parcs{\sa=.07em \sb=.03em
     \ifmmode \hbox{\rlap{.}}^{\scriptstyle\prime\kern -\sb\prime}\hbox{\kern -\sa}
     \else \rlap{.}$^{\scriptstyle\prime\kern -\sb\prime}$\kern -\sa\fi}
\def\parcm{\sa=.08em \sb=.03em
     \ifmmode \hbox{\rlap{.}\kern\sa}^{\scriptstyle\prime}\hbox{\kern-\sb}
     \else \rlap{.}\kern\sa$^{\scriptstyle\prime}$\kern-\sb\fi}
\def\ra[#1 #2 #3.#4]{#1\sup{h}#2\sup{m}#3\sup{s}\llap.#4}
\def\dec[#1 #2 #3.#4]{#1\deg#2\arcm#3\arcs\llap.#4}
\def\deco[#1 #2 #3]{#1\deg#2\arcm#3\arcs}
\def\rra[#1 #2]{#1\sup{h}#2\sup{m}}

\def\dots{\relax\ifmmode \ldots\else $\ldots$\fi}
\def\WHzsr{\ifmmode $W\,Hz\mo\,sr\mo$\else W\,Hz\mo\,sr\mo\fi}
\def\mHz{\ifmmode $\,mHz$\else \,mHz\fi}
\def\GHz{\ifmmode $\,GHz$\else \,GHz\fi}
\def\mKs{\ifmmode $\,mK\,s$^{1/2}\else \,mK\,s$^{1/2}$\fi}
\def\muKs{\ifmmode \,\mu$K\,s$^{1/2}\else \,$\mu$K\,s$^{1/2}$\fi}
\def\muKRJs{\ifmmode \,\mu$K$_{\rm RJ}$\,s$^{1/2}\else \,$\mu$K$_{\rm RJ}$\,s$^{1/2}$\fi}
\def\muKHz{\ifmmode \,\mu$K\,Hz$^{-1/2}\else \,$\mu$K\,Hz$^{-1/2}$\fi}
\def\MJysr{\ifmmode \,$MJy\,sr\mo$\else \,MJy\,sr\mo\fi}
\def\MJysrmK{\ifmmode \,$MJy\,sr\mo$\,mK$_{\rm CMB}\mo\else \,MJy\,sr\mo\,mK$_{\rm CMB}\mo$\fi}
\def\microns{\ifmmode \,\mu$m$\else \,$\mu$m\fi}

\def\muK{\ifmmode \,\mu$K$\else \,$\mu$\hbox{K}\fi}
\def\microK{\ifmmode \,\mu$K$\else \,$\mu$\hbox{K}\fi}
\def\muW{\ifmmode \,\mu$W$\else \,$\mu$\hbox{W}\fi}
\def\kms{\ifmmode $\,km\,s$^{-1}\else \,km\,s$^{-1}$\fi}
\def\kmsMpc{\ifmmode $\,\kms\,Mpc\mo$\else \,\kms\,Mpc\mo\fi}
\setsymbol{LFI:center:frequency:70GHz:units}{70.3\,GHz}
\setsymbol{LFI:center:frequency:44GHz:units}{44.1\,GHz}
\setsymbol{LFI:center:frequency:30GHz:units}{28.5\,GHz}

\setsymbol{LFI:center:frequency:70GHz}{70.3}
\setsymbol{LFI:center:frequency:44GHz}{44.1}
\setsymbol{LFI:center:frequency:30GHz}{28.5}

\setsymbol{LFI:center:frequency:LFI18:Rad:M:units}{71.7\GHz}
\setsymbol{LFI:center:frequency:LFI19:Rad:M:units}{67.5\GHz}
\setsymbol{LFI:center:frequency:LFI20:Rad:M:units}{69.2\GHz}
\setsymbol{LFI:center:frequency:LFI21:Rad:M:units}{70.4\GHz}
\setsymbol{LFI:center:frequency:LFI22:Rad:M:units}{71.5\GHz}
\setsymbol{LFI:center:frequency:LFI23:Rad:M:units}{70.8\GHz}
\setsymbol{LFI:center:frequency:LFI24:Rad:M:units}{44.4\GHz}
\setsymbol{LFI:center:frequency:LFI25:Rad:M:units}{44.0\GHz}
\setsymbol{LFI:center:frequency:LFI26:Rad:M:units}{43.9\GHz}
\setsymbol{LFI:center:frequency:LFI27:Rad:M:units}{28.3\GHz}
\setsymbol{LFI:center:frequency:LFI28:Rad:M:units}{28.8\GHz}
\setsymbol{LFI:center:frequency:LFI18:Rad:S:units}{70.1\GHz}
\setsymbol{LFI:center:frequency:LFI19:Rad:S:units}{69.6\GHz}
\setsymbol{LFI:center:frequency:LFI20:Rad:S:units}{69.5\GHz}
\setsymbol{LFI:center:frequency:LFI21:Rad:S:units}{69.5\GHz}
\setsymbol{LFI:center:frequency:LFI22:Rad:S:units}{72.8\GHz}
\setsymbol{LFI:center:frequency:LFI23:Rad:S:units}{71.3\GHz}
\setsymbol{LFI:center:frequency:LFI24:Rad:S:units}{44.1\GHz}
\setsymbol{LFI:center:frequency:LFI25:Rad:S:units}{44.1\GHz}
\setsymbol{LFI:center:frequency:LFI26:Rad:S:units}{44.1\GHz}
\setsymbol{LFI:center:frequency:LFI27:Rad:S:units}{28.5\GHz}
\setsymbol{LFI:center:frequency:LFI28:Rad:S:units}{28.2\GHz}

\setsymbol{LFI:center:frequency:LFI18:Rad:M}{71.7}
\setsymbol{LFI:center:frequency:LFI19:Rad:M}{67.5}
\setsymbol{LFI:center:frequency:LFI20:Rad:M}{69.2}
\setsymbol{LFI:center:frequency:LFI21:Rad:M}{70.4}
\setsymbol{LFI:center:frequency:LFI22:Rad:M}{71.5}
\setsymbol{LFI:center:frequency:LFI23:Rad:M}{70.8}
\setsymbol{LFI:center:frequency:LFI24:Rad:M}{44.4}
\setsymbol{LFI:center:frequency:LFI25:Rad:M}{44.0}
\setsymbol{LFI:center:frequency:LFI26:Rad:M}{43.9}
\setsymbol{LFI:center:frequency:LFI27:Rad:M}{28.3}
\setsymbol{LFI:center:frequency:LFI28:Rad:M}{28.8}
\setsymbol{LFI:center:frequency:LFI18:Rad:S}{70.1}
\setsymbol{LFI:center:frequency:LFI19:Rad:S}{69.6}
\setsymbol{LFI:center:frequency:LFI20:Rad:S}{69.5}
\setsymbol{LFI:center:frequency:LFI21:Rad:S}{69.5}
\setsymbol{LFI:center:frequency:LFI22:Rad:S}{72.8}
\setsymbol{LFI:center:frequency:LFI23:Rad:S}{71.3}
\setsymbol{LFI:center:frequency:LFI24:Rad:S}{44.1}
\setsymbol{LFI:center:frequency:LFI25:Rad:S}{44.1}
\setsymbol{LFI:center:frequency:LFI26:Rad:S}{44.1}
\setsymbol{LFI:center:frequency:LFI27:Rad:S}{28.5}
\setsymbol{LFI:center:frequency:LFI28:Rad:S}{28.2}

\setsymbol{LFI:white:noise:sensitivity:70GHz:units}{134.7\muKs}
\setsymbol{LFI:white:noise:sensitivity:44GHz:units}{164.7\muKs}
\setsymbol{LFI:white:noise:sensitivity:30GHz:units}{143.4\muKs}

\setsymbol{LFI:white:noise:sensitivity:70GHz}{134.7}
\setsymbol{LFI:white:noise:sensitivity:44GHz}{164.7}
\setsymbol{LFI:white:noise:sensitivity:30GHz}{143.4}

\setsymbol{LFI:white:noise:sensitivity:LFI18:Rad:M:units}{512.0\muKs}
\setsymbol{LFI:white:noise:sensitivity:LFI19:Rad:M:units}{581.4\muKs}
\setsymbol{LFI:white:noise:sensitivity:LFI20:Rad:M:units}{590.8\muKs}
\setsymbol{LFI:white:noise:sensitivity:LFI21:Rad:M:units}{455.2\muKs}
\setsymbol{LFI:white:noise:sensitivity:LFI22:Rad:M:units}{492.0\muKs}
\setsymbol{LFI:white:noise:sensitivity:LFI23:Rad:M:units}{507.7\muKs}
\setsymbol{LFI:white:noise:sensitivity:LFI24:Rad:M:units}{462.2\muKs}
\setsymbol{LFI:white:noise:sensitivity:LFI25:Rad:M:units}{413.6\muKs}
\setsymbol{LFI:white:noise:sensitivity:LFI26:Rad:M:units}{478.6\muKs}
\setsymbol{LFI:white:noise:sensitivity:LFI27:Rad:M:units}{277.7\muKs}
\setsymbol{LFI:white:noise:sensitivity:LFI28:Rad:M:units}{312.3\muKs}
\setsymbol{LFI:white:noise:sensitivity:LFI18:Rad:S:units}{465.7\muKs}
\setsymbol{LFI:white:noise:sensitivity:LFI19:Rad:S:units}{555.6\muKs}
\setsymbol{LFI:white:noise:sensitivity:LFI20:Rad:S:units}{623.2\muKs}
\setsymbol{LFI:white:noise:sensitivity:LFI21:Rad:S:units}{564.1\muKs}
\setsymbol{LFI:white:noise:sensitivity:LFI22:Rad:S:units}{534.4\muKs}
\setsymbol{LFI:white:noise:sensitivity:LFI23:Rad:S:units}{542.4\muKs}
\setsymbol{LFI:white:noise:sensitivity:LFI24:Rad:S:units}{399.2\muKs}
\setsymbol{LFI:white:noise:sensitivity:LFI25:Rad:S:units}{392.6\muKs}
\setsymbol{LFI:white:noise:sensitivity:LFI26:Rad:S:units}{418.6\muKs}
\setsymbol{LFI:white:noise:sensitivity:LFI27:Rad:S:units}{302.9\muKs}
\setsymbol{LFI:white:noise:sensitivity:LFI28:Rad:S:units}{285.3\muKs}

\setsymbol{LFI:white:noise:sensitivity:LFI18:Rad:M}{512.0}
\setsymbol{LFI:white:noise:sensitivity:LFI19:Rad:M}{581.4}
\setsymbol{LFI:white:noise:sensitivity:LFI20:Rad:M}{590.8}
\setsymbol{LFI:white:noise:sensitivity:LFI21:Rad:M}{455.2}
\setsymbol{LFI:white:noise:sensitivity:LFI22:Rad:M}{492.0}
\setsymbol{LFI:white:noise:sensitivity:LFI23:Rad:M}{507.7}
\setsymbol{LFI:white:noise:sensitivity:LFI24:Rad:M}{462.2}
\setsymbol{LFI:white:noise:sensitivity:LFI25:Rad:M}{413.6}
\setsymbol{LFI:white:noise:sensitivity:LFI26:Rad:M}{478.6}
\setsymbol{LFI:white:noise:sensitivity:LFI27:Rad:M}{277.7}
\setsymbol{LFI:white:noise:sensitivity:LFI28:Rad:M}{312.3}
\setsymbol{LFI:white:noise:sensitivity:LFI18:Rad:S}{465.7}
\setsymbol{LFI:white:noise:sensitivity:LFI19:Rad:S}{555.6}
\setsymbol{LFI:white:noise:sensitivity:LFI20:Rad:S}{623.2}
\setsymbol{LFI:white:noise:sensitivity:LFI21:Rad:S}{564.1}
\setsymbol{LFI:white:noise:sensitivity:LFI22:Rad:S}{534.4}
\setsymbol{LFI:white:noise:sensitivity:LFI23:Rad:S}{542.4}
\setsymbol{LFI:white:noise:sensitivity:LFI24:Rad:S}{399.2}
\setsymbol{LFI:white:noise:sensitivity:LFI25:Rad:S}{392.6}
\setsymbol{LFI:white:noise:sensitivity:LFI26:Rad:S}{418.6}
\setsymbol{LFI:white:noise:sensitivity:LFI27:Rad:S}{302.9}
\setsymbol{LFI:white:noise:sensitivity:LFI28:Rad:S}{285.3}

\setsymbol{LFI:knee:frequency:70GHz:units}{29.5\mHz}
\setsymbol{LFI:knee:frequency:44GHz:units}{56.2\mHz}
\setsymbol{LFI:knee:frequency:30GHz:units}{113.7\mHz}

\setsymbol{LFI:knee:frequency:70GHz}{29.5}
\setsymbol{LFI:knee:frequency:44GHz}{56.2}
\setsymbol{LFI:knee:frequency:30GHz}{113.7}

\setsymbol{LFI:knee:frequency:LFI18:Rad:M:units}{16.3\mHz}
\setsymbol{LFI:knee:frequency:LFI19:Rad:M:units}{15.1\mHz}
\setsymbol{LFI:knee:frequency:LFI20:Rad:M:units}{18.7\mHz}
\setsymbol{LFI:knee:frequency:LFI21:Rad:M:units}{37.2\mHz}
\setsymbol{LFI:knee:frequency:LFI22:Rad:M:units}{12.7\mHz}
\setsymbol{LFI:knee:frequency:LFI23:Rad:M:units}{34.6\mHz}
\setsymbol{LFI:knee:frequency:LFI24:Rad:M:units}{46.2\mHz}
\setsymbol{LFI:knee:frequency:LFI25:Rad:M:units}{24.9\mHz}
\setsymbol{LFI:knee:frequency:LFI26:Rad:M:units}{67.6\mHz}
\setsymbol{LFI:knee:frequency:LFI27:Rad:M:units}{187.4\mHz}
\setsymbol{LFI:knee:frequency:LFI28:Rad:M:units}{122.2\mHz}
\setsymbol{LFI:knee:frequency:LFI18:Rad:S:units}{17.7\mHz}
\setsymbol{LFI:knee:frequency:LFI19:Rad:S:units}{22.0\mHz}
\setsymbol{LFI:knee:frequency:LFI20:Rad:S:units}{8.7\mHz}
\setsymbol{LFI:knee:frequency:LFI21:Rad:S:units}{25.9\mHz}
\setsymbol{LFI:knee:frequency:LFI22:Rad:S:units}{15.8\mHz}
\setsymbol{LFI:knee:frequency:LFI23:Rad:S:units}{129.8\mHz}
\setsymbol{LFI:knee:frequency:LFI24:Rad:S:units}{100.9\mHz}
\setsymbol{LFI:knee:frequency:LFI25:Rad:S:units}{38.9\mHz}
\setsymbol{LFI:knee:frequency:LFI26:Rad:S:units}{58.9\mHz}
\setsymbol{LFI:knee:frequency:LFI27:Rad:S:units}{104.4\mHz}
\setsymbol{LFI:knee:frequency:LFI28:Rad:S:units}{40.7\mHz}

\setsymbol{LFI:knee:frequency:LFI18:Rad:M}{16.3}
\setsymbol{LFI:knee:frequency:LFI19:Rad:M}{15.1}
\setsymbol{LFI:knee:frequency:LFI20:Rad:M}{18.7}
\setsymbol{LFI:knee:frequency:LFI21:Rad:M}{37.2}
\setsymbol{LFI:knee:frequency:LFI22:Rad:M}{12.7}
\setsymbol{LFI:knee:frequency:LFI23:Rad:M}{34.6}
\setsymbol{LFI:knee:frequency:LFI24:Rad:M}{46.2}
\setsymbol{LFI:knee:frequency:LFI25:Rad:M}{24.9}
\setsymbol{LFI:knee:frequency:LFI26:Rad:M}{67.6}
\setsymbol{LFI:knee:frequency:LFI27:Rad:M}{187.4}
\setsymbol{LFI:knee:frequency:LFI28:Rad:M}{122.2}
\setsymbol{LFI:knee:frequency:LFI18:Rad:S}{17.7}
\setsymbol{LFI:knee:frequency:LFI19:Rad:S}{22.0}
\setsymbol{LFI:knee:frequency:LFI20:Rad:S}{8.7}
\setsymbol{LFI:knee:frequency:LFI21:Rad:S}{25.9}
\setsymbol{LFI:knee:frequency:LFI22:Rad:S}{15.8}
\setsymbol{LFI:knee:frequency:LFI23:Rad:S}{129.8}
\setsymbol{LFI:knee:frequency:LFI24:Rad:S}{100.9}
\setsymbol{LFI:knee:frequency:LFI25:Rad:S}{38.9}
\setsymbol{LFI:knee:frequency:LFI26:Rad:S}{58.9}
\setsymbol{LFI:knee:frequency:LFI27:Rad:S}{104.4}
\setsymbol{LFI:knee:frequency:LFI28:Rad:S}{40.7}

\setsymbol{LFI:slope:70GHz:units}{$-1.03$\mHz}
\setsymbol{LFI:slope:44GHz:units}{$-0.89$\mHz}
\setsymbol{LFI:slope:30GHz:units}{$-0.87$\mHz}

\setsymbol{LFI:slope:70GHz}{$-1.03$}
\setsymbol{LFI:slope:44GHz}{$-0.89$}
\setsymbol{LFI:slope:30GHz}{$-0.87$}

\setsymbol{LFI:slope:LFI18:Rad:M:units}{$-1.04$\mHz}
\setsymbol{LFI:slope:LFI19:Rad:M:units}{$-1.09$\mHz}
\setsymbol{LFI:slope:LFI20:Rad:M:units}{$-0.69$\mHz}
\setsymbol{LFI:slope:LFI21:Rad:M:units}{$-1.56$\mHz}
\setsymbol{LFI:slope:LFI22:Rad:M:units}{$-1.01$\mHz}
\setsymbol{LFI:slope:LFI23:Rad:M:units}{$-0.96$\mHz}
\setsymbol{LFI:slope:LFI24:Rad:M:units}{$-0.83$\mHz}
\setsymbol{LFI:slope:LFI25:Rad:M:units}{$-0.91$\mHz}
\setsymbol{LFI:slope:LFI26:Rad:M:units}{$-0.95$\mHz}
\setsymbol{LFI:slope:LFI27:Rad:M:units}{$-0.87$\mHz}
\setsymbol{LFI:slope:LFI28:Rad:M:units}{$-0.88$\mHz}
\setsymbol{LFI:slope:LFI18:Rad:S:units}{$-1.15$\mHz}
\setsymbol{LFI:slope:LFI19:Rad:S:units}{$-1.00$\mHz}
\setsymbol{LFI:slope:LFI20:Rad:S:units}{$-0.95$\mHz}
\setsymbol{LFI:slope:LFI21:Rad:S:units}{$-0.92$\mHz}
\setsymbol{LFI:slope:LFI22:Rad:S:units}{$-1.01$\mHz}
\setsymbol{LFI:slope:LFI23:Rad:S:units}{$-0.95$\mHz}
\setsymbol{LFI:slope:LFI24:Rad:S:units}{$-0.73$\mHz}
\setsymbol{LFI:slope:LFI25:Rad:S:units}{$-1.16$\mHz}
\setsymbol{LFI:slope:LFI26:Rad:S:units}{$-0.79$\mHz}
\setsymbol{LFI:slope:LFI27:Rad:S:units}{$-0.82$\mHz}
\setsymbol{LFI:slope:LFI28:Rad:S:units}{$-0.91$\mHz}

\setsymbol{LFI:slope:LFI18:Rad:M}{$-1.04$}
\setsymbol{LFI:slope:LFI19:Rad:M}{$-1.09$}
\setsymbol{LFI:slope:LFI20:Rad:M}{$-0.69$}
\setsymbol{LFI:slope:LFI21:Rad:M}{$-1.56$}
\setsymbol{LFI:slope:LFI22:Rad:M}{$-1.01$}
\setsymbol{LFI:slope:LFI23:Rad:M}{$-0.96$}
\setsymbol{LFI:slope:LFI24:Rad:M}{$-0.83$}
\setsymbol{LFI:slope:LFI25:Rad:M}{$-0.91$}
\setsymbol{LFI:slope:LFI26:Rad:M}{$-0.95$}
\setsymbol{LFI:slope:LFI27:Rad:M}{$-0.87$}
\setsymbol{LFI:slope:LFI28:Rad:M}{$-0.88$}
\setsymbol{LFI:slope:LFI18:Rad:S}{$-1.15$}
\setsymbol{LFI:slope:LFI19:Rad:S}{$-1.00$}
\setsymbol{LFI:slope:LFI20:Rad:S}{$-0.95$}
\setsymbol{LFI:slope:LFI21:Rad:S}{$-0.92$}
\setsymbol{LFI:slope:LFI22:Rad:S}{$-1.01$}
\setsymbol{LFI:slope:LFI23:Rad:S}{$-0.95$}
\setsymbol{LFI:slope:LFI24:Rad:S}{$-0.73$}
\setsymbol{LFI:slope:LFI25:Rad:S}{$-1.16$}
\setsymbol{LFI:slope:LFI26:Rad:S}{$-0.79$}
\setsymbol{LFI:slope:LFI27:Rad:S}{$-0.82$}
\setsymbol{LFI:slope:LFI28:Rad:S}{$-0.91$}

\setsymbol{LFI:FWHM:70GHz:units}{13\parcm01}
\setsymbol{LFI:FWHM:44GHz:units}{27\parcm92}
\setsymbol{LFI:FWHM:30GHz:units}{32\parcm65}

\setsymbol{LFI:FWHM:70GHz}{13.01}
\setsymbol{LFI:FWHM:44GHz}{27.92}
\setsymbol{LFI:FWHM:30GHz}{32.65}

\setsymbol{LFI:FWHM:LFI18:units}{13\parcm39}
\setsymbol{LFI:FWHM:LFI19:units}{13\parcm01}
\setsymbol{LFI:FWHM:LFI20:units}{12\parcm75}
\setsymbol{LFI:FWHM:LFI21:units}{12\parcm74}
\setsymbol{LFI:FWHM:LFI22:units}{12\parcm87}
\setsymbol{LFI:FWHM:LFI23:units}{13\parcm27}
\setsymbol{LFI:FWHM:LFI24:units}{22\parcm98}
\setsymbol{LFI:FWHM:LFI25:units}{30\parcm46}
\setsymbol{LFI:FWHM:LFI26:units}{30\parcm31}
\setsymbol{LFI:FWHM:LFI27:units}{32\parcm65}
\setsymbol{LFI:FWHM:LFI28:units}{32\parcm66}

\setsymbol{LFI:FWHM:LFI18}{13.39}
\setsymbol{LFI:FWHM:LFI19}{13.01}
\setsymbol{LFI:FWHM:LFI20}{12.75}
\setsymbol{LFI:FWHM:LFI21}{12.74}
\setsymbol{LFI:FWHM:LFI22}{12.87}
\setsymbol{LFI:FWHM:LFI23}{13.27}
\setsymbol{LFI:FWHM:LFI24}{22.98}
\setsymbol{LFI:FWHM:LFI25}{30.46}
\setsymbol{LFI:FWHM:LFI26}{30.31}
\setsymbol{LFI:FWHM:LFI27}{32.65}
\setsymbol{LFI:FWHM:LFI28}{32.66}

\setsymbol{LFI:FWHM:uncertainty:LFI18:units}{0.170\arcm}
\setsymbol{LFI:FWHM:uncertainty:LFI19:units}{0.174\arcm}
\setsymbol{LFI:FWHM:uncertainty:LFI20:units}{0.170\arcm}
\setsymbol{LFI:FWHM:uncertainty:LFI21:units}{0.156\arcm}
\setsymbol{LFI:FWHM:uncertainty:LFI22:units}{0.164\arcm}
\setsymbol{LFI:FWHM:uncertainty:LFI23:units}{0.171\arcm}
\setsymbol{LFI:FWHM:uncertainty:LFI24:units}{0.652\arcm}
\setsymbol{LFI:FWHM:uncertainty:LFI25:units}{1.075\arcm}
\setsymbol{LFI:FWHM:uncertainty:LFI26:units}{1.131\arcm}
\setsymbol{LFI:FWHM:uncertainty:LFI27:units}{1.266\arcm}
\setsymbol{LFI:FWHM:uncertainty:LFI28:units}{1.287\arcm}

\setsymbol{LFI:FWHM:uncertainty:LFI18}{0.170}
\setsymbol{LFI:FWHM:uncertainty:LFI19}{0.174}
\setsymbol{LFI:FWHM:uncertainty:LFI20}{0.170}
\setsymbol{LFI:FWHM:uncertainty:LFI21}{0.156}
\setsymbol{LFI:FWHM:uncertainty:LFI22}{0.164}
\setsymbol{LFI:FWHM:uncertainty:LFI23}{0.171}
\setsymbol{LFI:FWHM:uncertainty:LFI24}{0.652}
\setsymbol{LFI:FWHM:uncertainty:LFI25}{1.075}
\setsymbol{LFI:FWHM:uncertainty:LFI26}{1.131}
\setsymbol{LFI:FWHM:uncertainty:LFI27}{1.266}
\setsymbol{LFI:FWHM:uncertainty:LFI28}{1.287}

\setsymbol{HFI:center:frequency:100GHz:units}{100\,GHz}
\setsymbol{HFI:center:frequency:143GHz:units}{143\,GHz}
\setsymbol{HFI:center:frequency:217GHz:units}{217\,GHz}
\setsymbol{HFI:center:frequency:353GHz:units}{353\,GHz}
\setsymbol{HFI:center:frequency:545GHz:units}{545\,GHz}
\setsymbol{HFI:center:frequency:857GHz:units}{857\,GHz}

\setsymbol{HFI:center:frequency:100GHz}{100}
\setsymbol{HFI:center:frequency:143GHz}{143}
\setsymbol{HFI:center:frequency:217GHz}{217}
\setsymbol{HFI:center:frequency:353GHz}{353}
\setsymbol{HFI:center:frequency:545GHz}{545}
\setsymbol{HFI:center:frequency:857GHz}{857}

\setsymbol{HFI:Ndetectors:100GHz}{8}
\setsymbol{HFI:Ndetectors:143GHz}{11}
\setsymbol{HFI:Ndetectors:217GHz}{12}
\setsymbol{HFI:Ndetectors:353GHz}{12}
\setsymbol{HFI:Ndetectors:545GHz}{3}
\setsymbol{HFI:Ndetectors:857GHz}{4}

\setsymbol{HFI:FWHM:Maps:100GHz:units}{9\parcm88}
\setsymbol{HFI:FWHM:Maps:143GHz:units}{7\parcm18}
\setsymbol{HFI:FWHM:Maps:217GHz:units}{4\parcm87}
\setsymbol{HFI:FWHM:Maps:353GHz:units}{4\parcm65}
\setsymbol{HFI:FWHM:Maps:545GHz:units}{4\parcm72}
\setsymbol{HFI:FWHM:Maps:857GHz:units}{4\parcm39}
\setsymbol{HFI:FWHM:Maps:100GHz}{9.88}
\setsymbol{HFI:FWHM:Maps:143GHz}{7.18}
\setsymbol{HFI:FWHM:Maps:217GHz}{4.87}
\setsymbol{HFI:FWHM:Maps:353GHz}{4.65}
\setsymbol{HFI:FWHM:Maps:545GHz}{4.72}
\setsymbol{HFI:FWHM:Maps:857GHz}{4.39}

\setsymbol{HFI:beam:ellipticity:Maps:100GHz}{1.15}
\setsymbol{HFI:beam:ellipticity:Maps:143GHz}{1.01}
\setsymbol{HFI:beam:ellipticity:Maps:217GHz}{1.06}
\setsymbol{HFI:beam:ellipticity:Maps:353GHz}{1.05}
\setsymbol{HFI:beam:ellipticity:Maps:545GHz}{1.14}
\setsymbol{HFI:beam:ellipticity:Maps:857GHz}{1.19}

\setsymbol{HFI:FWHM:Mars:100GHz:units}{9\parcm37}
\setsymbol{HFI:FWHM:Mars:143GHz:units}{7\parcm04}
\setsymbol{HFI:FWHM:Mars:217GHz:units}{4\parcm68}
\setsymbol{HFI:FWHM:Mars:353GHz:units}{4\parcm43}
\setsymbol{HFI:FWHM:Mars:545GHz:units}{3\parcm80}
\setsymbol{HFI:FWHM:Mars:857GHz:units}{3\parcm67}

\setsymbol{HFI:FWHM:Mars:100GHz}{9.37}
\setsymbol{HFI:FWHM:Mars:143GHz}{7.04}
\setsymbol{HFI:FWHM:Mars:217GHz}{4.68}
\setsymbol{HFI:FWHM:Mars:353GHz}{4.43}
\setsymbol{HFI:FWHM:Mars:545GHz}{3.80}
\setsymbol{HFI:FWHM:Mars:857GHz}{3.67}

\setsymbol{HFI:beam:ellipticity:Mars:100GHz}{1.18}
\setsymbol{HFI:beam:ellipticity:Mars:143GHz}{1.03}
\setsymbol{HFI:beam:ellipticity:Mars:217GHz}{1.14}
\setsymbol{HFI:beam:ellipticity:Mars:353GHz}{1.09}
\setsymbol{HFI:beam:ellipticity:Mars:545GHz}{1.25}
\setsymbol{HFI:beam:ellipticity:Mars:857GHz}{1.03}

\setsymbol{HFI:CMB:relative:calibration:100GHz}{$\lsim 1\%$}
\setsymbol{HFI:CMB:relative:calibration:143GHz}{$\lsim 1\%$}
\setsymbol{HFI:CMB:relative:calibration:217GHz}{$\lsim 1\%$}
\setsymbol{HFI:CMB:relative:calibration:353GHz}{$\lsim 1\%$}
\setsymbol{HFI:CMB:relative:calibration:545GHz}{}
\setsymbol{HFI:CMB:relative:calibration:857GHz}{}

\setsymbol{HFI:CMB:absolute:calibration:100GHz}{$\lsim 2\%$}
\setsymbol{HFI:CMB:absolute:calibration:143GHz}{$\lsim 2\%$}
\setsymbol{HFI:CMB:absolute:calibration:217GHz}{$\lsim 2\%$}
\setsymbol{HFI:CMB:absolute:calibration:353GHz}{$\lsim 2\%$}
\setsymbol{HFI:CMB:absolute:calibration:545GHz}{}
\setsymbol{HFI:CMB:absolute:calibration:857GHz}{}

\setsymbol{HFI:FIRAS:gain:calibration:accuracy:statistical:100GHz}{}
\setsymbol{HFI:FIRAS:gain:calibration:accuracy:statistical:143GHz}{}
\setsymbol{HFI:FIRAS:gain:calibration:accuracy:statistical:217GHz}{}
\setsymbol{HFI:FIRAS:gain:calibration:accuracy:statistical:353GHz}{2.5\%}
\setsymbol{HFI:FIRAS:gain:calibration:accuracy:statistical:545GHz}{1\%}
\setsymbol{HFI:FIRAS:gain:calibration:accuracy:statistical:857GHz}{0.5\%}

\setsymbol{HFI:FIRAS:gain:calibration:accuracy:systematic:100GHz}{}
\setsymbol{HFI:FIRAS:gain:calibration:accuracy:systematic:143GHz}{}
\setsymbol{HFI:FIRAS:gain:calibration:accuracy:systematic:217GHz}{}
\setsymbol{HFI:FIRAS:gain:calibration:accuracy:systematic:353GHz}{}
\setsymbol{HFI:FIRAS:gain:calibration:accuracy:systematic:545GHz}{7\%}
\setsymbol{HFI:FIRAS:gain:calibration:accuracy:systematic:857GHz}{7\%}

\setsymbol{HFI:FIRAS:zero:point:accuracy:100GHz:units}{0.8\MJysr}
\setsymbol{HFI:FIRAS:zero:point:accuracy:143GHz:units}{}
\setsymbol{HFI:FIRAS:zero:point:accuracy:217GHz:units}{}
\setsymbol{HFI:FIRAS:zero:point:accuracy:353GHz:units}{1.4\MJysr}
\setsymbol{HFI:FIRAS:zero:point:accuracy:545GHz:units}{2.2\MJysr}
\setsymbol{HFI:FIRAS:zero:point:accuracy:857GHz:units}{1.7\MJysr}

\setsymbol{HFI:FIRAS:zero:point:accuracy:100GHz}{0.8}
\setsymbol{HFI:FIRAS:zero:point:accuracy:143GHz}{}
\setsymbol{HFI:FIRAS:zero:point:accuracy:217GHz}{}
\setsymbol{HFI:FIRAS:zero:point:accuracy:353GHz}{1.4}
\setsymbol{HFI:FIRAS:zero:point:accuracy:545GHz}{2.2}
\setsymbol{HFI:FIRAS:zero:point:accuracy:857GHz}{1.7}

\setsymbol{HFI:unit:conversion:100GHz:units}{0.2415\MJysrmK}
\setsymbol{HFI:unit:conversion:143GHz:units}{0.3694\MJysrmK}
\setsymbol{HFI:unit:conversion:217GHz:units}{0.4811\MJysrmK}
\setsymbol{HFI:unit:conversion:353GHz:units}{0.2883\MJysrmK}
\setsymbol{HFI:unit:conversion:545GHz:units}{0.05826\MJysrmK}
\setsymbol{HFI:unit:conversion:857GHz:units}{0.002238\MJysrmK}

\setsymbol{HFI:unit:conversion:100GHz}{0.2415}
\setsymbol{HFI:unit:conversion:143GHz}{0.3694}
\setsymbol{HFI:unit:conversion:217GHz}{0.4811}
\setsymbol{HFI:unit:conversion:353GHz}{0.2883}
\setsymbol{HFI:unit:conversion:545GHz}{0.05826}
\setsymbol{HFI:unit:conversion:857GHz}{0.002238}

\setsymbol{HFI:colour:correction:alpha=-2:V1.01:100GHz}{0.9893}
\setsymbol{HFI:colour:correction:alpha=-2:V1.01:143GHz}{0.9759}
\setsymbol{HFI:colour:correction:alpha=-2:V1.01:217GHz}{1.0007}
\setsymbol{HFI:colour:correction:alpha=-2:V1.01:353GHz}{1.0028}
\setsymbol{HFI:colour:correction:alpha=-2:V1.01:545GHz}{1.0019}
\setsymbol{HFI:colour:correction:alpha=-2:V1.01:857GHz}{0.9889}

\setsymbol{HFI:colour:correction:alpha=0:V1.01:100GHz}{1.0008}
\setsymbol{HFI:colour:correction:alpha=0:V1.01:143GHz}{1.0148}
\setsymbol{HFI:colour:correction:alpha=0:V1.01:217GHz}{0.9909}
\setsymbol{HFI:colour:correction:alpha=0:V1.01:353GHz}{0.9888}
\setsymbol{HFI:colour:correction:alpha=0:V1.01:545GHz}{0.9878}
\setsymbol{HFI:colour:correction:alpha=0:V1.01:857GHz}{1.0014}

\providecommand{\sorthelp}[1]{}

%% file: PIP_105_Hernandez_authors_and_institutes.tex
\author{\small
Planck Collaboration: P.~A.~R.~Ade\inst{82}
\and
N.~Aghanim\inst{55}
\and
M.~Arnaud\inst{70}
\and
M.~Ashdown\inst{66, 6}
\and
E.~Aubourg\inst{1}
\and
J.~Aumont\inst{55}
\and
C.~Baccigalupi\inst{81}
\and
A.~J.~Banday\inst{89, 10}
\and
R.~B.~Barreiro\inst{61}
\and
N.~Bartolo\inst{28, 62}
% ...
}
\institute{\small
APC, AstroParticule et Cosmologie, Universit\'{e} Paris Diderot, CNRS/IN2P3, CEA/lrfu, Observatoire de Paris, Sorbonne Paris Cit\'{e}, 10, rue Alice Domon et L\'{e}onie Duquet, 75205 Paris Cedex 13, France\goodbreak
\and
Aalto University Mets\"{a}hovi Radio Observatory, P.O. Box 13000, FI-00076 AALTO, Finland\goodbreak
\and
African Institute for Mathematical Sciences, 6-8 Melrose Road, Muizenberg, Cape Town, South Africa\goodbreak
\and
Agenzia Spaziale Italiana Science Data Center, Via del Politecnico snc, 00133, Roma, Italy\goodbreak
\and
Aix Marseille Universit\'{e}, CNRS, LAM (Laboratoire d'Astrophysique de Marseille) UMR 7326, 13388, Marseille, France\goodbreak
\and
Astrophysics Group, Cavendish Laboratory, University of Cambridge, J J Thomson Avenue, Cambridge CB3 0HE, U.K.\goodbreak
\and
Astrophysics \& Cosmology Research Unit, School of Mathematics, Statistics \& Computer Science, University of KwaZulu-Natal, Westville Campus, Private Bag X54001, Durban 4000, South Africa\goodbreak
\and
Atacama Large Millimeter/submillimeter Array, ALMA Santiago Central Offices, Alonso de Cordova 3107, Vitacura, Casilla 763 0355, Santiago, Chile\goodbreak
\and
CITA, University of Toronto, 60 St. George St., Toronto, ON M5S 3H8, Canada\goodbreak
\and
CNRS, IRAP, 9 Av. colonel Roche, BP 44346, F-31028 Toulouse cedex 4, France\goodbreak
\and
California Institute of Technology, Pasadena, California, U.S.A.\goodbreak
\and
Centro de Estudios de F\'{i}sica del Cosmos de Arag\'{o}n (CEFCA), Plaza San Juan, 1, planta 2, E-44001, Teruel, Spain\goodbreak
\and
Computational Cosmology Center, Lawrence Berkeley National Laboratory, Berkeley, California, U.S.A.\goodbreak
\and
DSM/Irfu/SPP, CEA-Saclay, F-91191 Gif-sur-Yvette Cedex, France\goodbreak
\and
DTU Space, National Space Institute, Technical University of Denmark, Elektrovej 327, DK-2800 Kgs. Lyngby, Denmark\goodbreak
\and
D\'{e}partement de Physique Th\'{e}orique, Universit\'{e} de Gen\`{e}ve, 24, Quai E. Ansermet,1211 Gen\`{e}ve 4, Switzerland\goodbreak
\and
Departamento de Astrof\'{i}sica, Universidad de La Laguna (ULL), E-38206 La Laguna, Tenerife, Spain\goodbreak
\and
Departamento de F\'{\i}sica, Universidad de Oviedo, Avda. Calvo Sotelo s/n, Oviedo, Spain\goodbreak
\and
Department of Astrophysics/IMAPP, Radboud University Nijmegen, P.O. Box 9010, 6500 GL Nijmegen, The Netherlands\goodbreak
\and
Department of Physics \& Astronomy, University of British Columbia, 6224 Agricultural Road, Vancouver, British Columbia, Canada\goodbreak
\and
Department of Physics and Astronomy, Dana and David Dornsife College of Letter, Arts and Sciences, University of Southern California, Los Angeles, CA 90089, U.S.A.\goodbreak
\and
Department of Physics and Astronomy, University College London, London WC1E 6BT, U.K.\goodbreak
\and
Department of Physics, Florida State University, Keen Physics Building, 77 Chieftan Way, Tallahassee, Florida, U.S.A.\goodbreak
\and
Department of Physics, Gustaf H\"{a}llstr\"{o}min katu 2a, University of Helsinki, Helsinki, Finland\goodbreak
\and
Department of Physics, Princeton University, Princeton, New Jersey, U.S.A.\goodbreak
\and
Department of Physics, University of California, Santa Barbara, California, U.S.A.\goodbreak
\and
Department of Physics, University of Illinois at Urbana-Champaign, 1110 West Green Street, Urbana, Illinois, U.S.A.\goodbreak
\and
Dipartimento di Fisica e Astronomia G. Galilei, Universit\`{a} degli Studi di Padova, via Marzolo 8, 35131 Padova, Italy\goodbreak
\and
Dipartimento di Fisica e Scienze della Terra, Universit\`{a} di Ferrara, Via Saragat 1, 44122 Ferrara, Italy\goodbreak
\and
Dipartimento di Fisica, Universit\`{a} La Sapienza, P. le A. Moro 2, Roma, Italy\goodbreak
\and
Dipartimento di Fisica, Universit\`{a} degli Studi di Milano, Via Celoria, 16, Milano, Italy\goodbreak
\and
Dipartimento di Fisica, Universit\`{a} degli Studi di Trieste, via A. Valerio 2, Trieste, Italy\goodbreak
\and
Dipartimento di Matematica, Universit\`{a} di Roma Tor Vergata, Via della Ricerca Scientifica, 1, Roma, Italy\goodbreak
\and
Discovery Center, Niels Bohr Institute, Blegdamsvej 17, Copenhagen, Denmark\goodbreak
\and
European Southern Observatory, ESO Vitacura, Alonso de Cordova 3107, Vitacura, Casilla 19001, Santiago, Chile\goodbreak
\and
European Space Agency, ESAC, Planck Science Office, Camino bajo del Castillo, s/n, Urbanizaci\'{o}n Villafranca del Castillo, Villanueva de la Ca\~{n}ada, Madrid, Spain\goodbreak
\and
European Space Agency, ESTEC, Keplerlaan 1, 2201 AZ Noordwijk, The Netherlands\goodbreak
\and
Facolt\`{a} di Ingegneria, Universit\`{a} degli Studi e-Campus, Via Isimbardi 10, Novedrate (CO), 22060, Italy\goodbreak
\and
Finnish Centre for Astronomy with ESO (FINCA), University of Turku, V\"{a}is\"{a}l\"{a}ntie 20, FIN-21500, Piikki\"{o}, Finland\goodbreak
\and
Gran Sasso Science Institute, INFN, viale F. Crispi 7, 67100 L'Aquila, Italy\goodbreak
\and
HGSFP and University of Heidelberg, Theoretical Physics Department, Philosophenweg 16, 69120, Heidelberg, Germany\goodbreak
\and
Helsinki Institute of Physics, Gustaf H\"{a}llstr\"{o}min katu 2, University of Helsinki, Helsinki, Finland\goodbreak
\and
INAF - Osservatorio Astronomico di Padova, Vicolo dell'Osservatorio 5, Padova, Italy\goodbreak
\and
INAF - Osservatorio Astronomico di Roma, via di Frascati 33, Monte Porzio Catone, Italy\goodbreak
\and
INAF - Osservatorio Astronomico di Trieste, Via G.B. Tiepolo 11, Trieste, Italy\goodbreak
\and
INAF/IASF Bologna, Via Gobetti 101, Bologna, Italy\goodbreak
\and
INAF/IASF Milano, Via E. Bassini 15, Milano, Italy\goodbreak
\and
INFN, Sezione di Bologna, Via Irnerio 46, I-40126, Bologna, Italy\goodbreak
\and
INFN, Sezione di Roma 1, Universit\`{a} di Roma Sapienza, Piazzale Aldo Moro 2, 00185, Roma, Italy\goodbreak
\and
INFN, Sezione di Roma 2, Universit\`{a} di Roma Tor Vergata, Via della Ricerca Scientifica, 1, Roma, Italy\goodbreak
\and
INFN/National Institute for Nuclear Physics, Via Valerio 2, I-34127 Trieste, Italy\goodbreak
\and
IPAG: Institut de Plan\'{e}tologie et d'Astrophysique de Grenoble, Universit\'{e} Grenoble Alpes, IPAG, F-38000 Grenoble, France, CNRS, IPAG, F-38000 Grenoble, France\goodbreak
\and
Imperial College London, Astrophysics group, Blackett Laboratory, Prince Consort Road, London, SW7 2AZ, U.K.\goodbreak
\and
Infrared Processing and Analysis Center, California Institute of Technology, Pasadena, CA 91125, U.S.A.\goodbreak
\and
Institut d'Astrophysique Spatiale, CNRS (UMR8617) Universit\'{e} Paris-Sud 11, B\^{a}timent 121, Orsay, France\goodbreak
\and
Institut d'Astrophysique de Paris, CNRS (UMR7095), 98 bis Boulevard Arago, F-75014, Paris, France\goodbreak
\and
Institute of Astronomy, University of Cambridge, Madingley Road, Cambridge CB3 0HA, U.K.\goodbreak
\and
Institute of Theoretical Astrophysics, University of Oslo, Blindern, Oslo, Norway\goodbreak
\and
Instituto Nacional de Astrof\'{\i}sica, \'{O}ptica y Electr\'{o}nica (INAOE), Apartado Postal 51 y 216, 72000 Puebla, M\'{e}xico\goodbreak
\and
Instituto de Astrof\'{\i}sica de Canarias, C/V\'{\i}a L\'{a}ctea s/n, La Laguna, Tenerife, Spain\goodbreak
\and
Instituto de F\'{\i}sica de Cantabria (CSIC-Universidad de Cantabria), Avda. de los Castros s/n, Santander, Spain\goodbreak
\and
Istituto Nazionale di Fisica Nucleare, Sezione di Padova, via Marzolo 8, I-35131 Padova, Italy\goodbreak
\and
Jet Propulsion Laboratory, California Institute of Technology, 4800 Oak Grove Drive, Pasadena, California, U.S.A.\goodbreak
\and
Jodrell Bank Centre for Astrophysics, Alan Turing Building, School of Physics and Astronomy, The University of Manchester, Oxford Road, Manchester, M13 9PL, U.K.\goodbreak
\and
Kavli Institute for Cosmological Physics, University of Chicago, Chicago, IL 60637, USA\goodbreak
\and
Kavli Institute for Cosmology Cambridge, Madingley Road, Cambridge, CB3 0HA, U.K.\goodbreak
\and
Kazan Federal University, 18 Kremlyovskaya St., Kazan, 420008, Russia\goodbreak
\and
LAL, Universit\'{e} Paris-Sud, CNRS/IN2P3, Orsay, France\goodbreak
\and
LERMA, CNRS, Observatoire de Paris, 61 Avenue de l'Observatoire, Paris, France\goodbreak
\and
Laboratoire AIM, IRFU/Service d'Astrophysique - CEA/DSM - CNRS - Universit\'{e} Paris Diderot, B\^{a}t. 709, CEA-Saclay, F-91191 Gif-sur-Yvette Cedex, France\goodbreak
\and
Laboratoire Traitement et Communication de l'Information, CNRS (UMR 5141) and T\'{e}l\'{e}com ParisTech, 46 rue Barrault F-75634 Paris Cedex 13, France\goodbreak
\and
Laboratoire de Physique Subatomique et Cosmologie, Universit\'{e} Grenoble-Alpes, CNRS/IN2P3, 53, rue des Martyrs, 38026 Grenoble Cedex, France\goodbreak
\and
Laboratoire de Physique Th\'{e}orique, Universit\'{e} Paris-Sud 11 \& CNRS, B\^{a}timent 210, 91405 Orsay, France\goodbreak
\and
Lebedev Physical Institute of the Russian Academy of Sciences, Astro Space Centre, 84/32 Profsoyuznaya st., Moscow, GSP-7, 117997, Russia\goodbreak
\and
Leibniz-Institut f\"{u}r Astrophysik Potsdam (AIP), An der Sternwarte 16, D-14482 Potsdam, Germany\goodbreak
\and
Max-Planck-Institut f\"{u}r Astrophysik, Karl-Schwarzschild-Str. 1, 85741 Garching, Germany\goodbreak
\and
National University of Ireland, Department of Experimental Physics, Maynooth, Co. Kildare, Ireland\goodbreak
\and
Nicolaus Copernicus Astronomical Center, Bartycka 18, 00-716 Warsaw, Poland\goodbreak
\and
Niels Bohr Institute, Blegdamsvej 17, Copenhagen, Denmark\goodbreak
\and
Optical Science Laboratory, University College London, Gower Street, London, U.K.\goodbreak
\and
SISSA, Astrophysics Sector, via Bonomea 265, 34136, Trieste, Italy\goodbreak
\and
School of Physics and Astronomy, Cardiff University, Queens Buildings, The Parade, Cardiff, CF24 3AA, U.K.\goodbreak
\and
Sorbonne Universit\'{e}-UPMC, UMR7095, Institut d'Astrophysique de Paris, 98 bis Boulevard Arago, F-75014, Paris, France\goodbreak
\and
Space Research Institute (IKI), Russian Academy of Sciences, Profsoyuznaya Str, 84/32, Moscow, 117997, Russia\goodbreak
\and
Space Sciences Laboratory, University of California, Berkeley, California, U.S.A.\goodbreak
\and
Special Astrophysical Observatory, Russian Academy of Sciences, Nizhnij Arkhyz, Zelenchukskiy region, Karachai-Cherkessian Republic, 369167, Russia\goodbreak
\and
Sub-Department of Astrophysics, University of Oxford, Keble Road, Oxford OX1 3RH, U.K.\goodbreak
\and
UPMC Univ Paris 06, UMR7095, 98 bis Boulevard Arago, F-75014, Paris, France\goodbreak
\and
Universit\'{e} de Toulouse, UPS-OMP, IRAP, F-31028 Toulouse cedex 4, France\goodbreak
% ...
}